# Mechanisms for covalent immobilization of horseradish peroxidase on ion beam treated polyethylene

Alexey V. Kondyurin, Pourandokht Naseri, Jennifer M. R. Tilley, Neil J. Nosworthy, Marcela M.M. Bilek and David R. McKenzie

Applied and Plasma Physics, School of Physics (A28), University of Sydney, Sydney, NSW 2006, Australia

**Abstract**
The mechanism that provides the observed strong binding of biomolecules to polymer surfaces modified by ion beams is investigated. The surface of polyethylene (PE) was modified by plasma immersion ion implantation with nitrogen ions. Structure changes including carbonization and oxidation were observed in the modified surface layer of PE by Raman spectroscopy, FTIR ATR spectroscopy, atomic force microscopy, surface energy measurement and XPS spectroscopy. An observed high surface energy of the modified polyethylene was attributed to the presence of free radicals on the surface. The surface energy decay with storage time after PIII treatment was explained by a decay of the free radical concentration while the concentration of oxygen-containing groups increased with storage time. Horseradish peroxidase was covalently attached onto the modified PE surface. The enzymatic activity of covalently attached protein remained high. A mechanism based on the covalent attachment by the reaction of protein with free radicals in the modified surface is proposed. Appropriate blocking agents can block this reaction. All aminoacid residues can take part in the covalent attachment process, providing a universal mechanism of attachment for all proteins. The long-term activity of the modified layer to attach protein (at least 2 years) is explained by stabilisation of unpaired electrons in $sp^2$ carbon structures. The native conformation of attached protein is retained due to hydrophilic interactions in the interface region. A high concentration of free radicals on the surface can give multiple covalent bonds to the protein molecule and destroy the native conformation and with it the catalytic activity. The universal mechanism of protein attachment to free radicals could be extended to various methods of radiation damage of polymers.

## Introduction

The attachment of proteins to polymer surfaces provides a means of modifying the response of an organism to the surface and is therefore an important step for improving the biocompatibility and functionality of medical implantable devices [1]. In medicine, the immune response can cause adverse reactions to implanted prosthetics or during operations in which blood is exposed to the surfaces of medical devices such as heart-lung machines. In biosensors, attached proteins may be used to detect the presence of molecules in the environment. The attached protein must be strongly bound to the surface to prevent it being washed off under operational conditions including a high rate of flow of liquid across the surface [2]. Additionally the surface must allow the protein to preserve its bioactivity [3].

Given the rigorous protocols that must be undertaken to obtain approvals for the use of new polymer materials in medical applications, it is preferable to modify the surface of an existing polymer than to develop an entirely new polymer material. The preparation of polymer surfaces for protein binding can be done using a number of chemical and physical modifications, such as the attachment of linker molecules to provide covalent binding through specific active groups [4, 5]; plasma treatment [6-12]; UV treatment [13-16] and ion beam implantation [17-20].

Despite its importance in applications, the attachment mechanism of proteins on polymer surfaces is not yet well understood [21-27]. The uncertainty of the mechanism could make the behaviour of the surface unpredictable in a given application. In a limited number of applications, the physical adsorption on a polymer surface is acceptable. However, in some applications, the stronger attachment provided by covalent bonding is desirable and linker groups have been employed to achieve this [28]. The mechanism of attachment provided by linker groups is usually assumed to be covalent bonding via the chemistry that is expected for the active group of the linker and the protein, based on a general knowledge of polymer and protein chemistry. However, without clear experimental evidence of the reaction products and without analysis of the reaction conditions, it should not be assumed that the expected chemistry has in fact occurred. The linker binding method has disadvantages as it restricts significantly the range of available substrates, restricting its use in biomedical applications, which usually need to satisfy specific requirements. Biomechanical properties, low toxicity, biostability, suitability for sterilization are all important and may dictate the choice of substrate. Also, the linker method is not suitable for all proteins since a suitable group is required in the external protein surface and the linker chemistry could affect conformation.

The methods of achieving protein binding based on radiation damage in polymers (plasma treatment, UV light and ion beam implantation) can provide strong binding, possibly covalent binding, to polymer surfaces and has been achieved without the use of specific linker groups [29-35]. These methods appear to be widely applicable to many polymer surfaces and to many proteins. The mechanism of binding has been assumed to be through physical intermolecular interactions, morphology of the surface, reactive groups, including oxygen containing groups or radical containing groups, but there is at present, no direct evidence as to the actual mechanism.

When a polymer is subjected to modification by an ion beam, energy transfer from the impacting ions to the target atoms occurs, generating broken bonds, ionisation and electronic and vibrational excitation of the macromolecules. A large number of investigations of different kinds of polymers under ion beam impact have been reported and overviewed [36-38]. The ion beam implantation for improvement of protein attachment on polyethylene [39], polytetrafluorethylene [40], polystyrene [41] and Nylon [42] has been reported. Strong bonding between the modified polymer surface and active enzyme molecules including horseradish peroxidase (HRP), soybean peroxidase, catalase and tropoelastin was observed together with the preservation of the enzyme's catalytic activity.

The proposed mechanisms of protein attachment that have been proposed for binding to modified polymers are in contradiction with each other and do not explain observations. For example, an appearance of oxygen containing groups in polymer surface after radiation damage causes an increase in surface energy, especially its polar part [10, 36-38, 43-48]. If the mechanism is of protein binding is assumed to be via physical adsorption, protein attachment should be stronger on non-polar surfaces [49, 50]. Therefore, the modified surface should be less attractive to protein, a result that is not supported by experimental results [29-34]. The increase of protein attachment after modification has been associated with an increase of the effective surface area, but in a number of investigations of polymers after radiation modifica-tion the surface morphology does not change or the surface becomes even smoother [51, 52]. The appearance of oxygen containing groups has been proposed as the reason for strong protein attachment [53], but polymers with oxygen containing groups in macromolecules do not provide strong attachment of proteins without modification [54]. The present investigation aims to determine the actual mechanism of protein attachment realised one example of polymer surface modification - ion beam implantation achieved by plasma immersion ion implantation (PIII). In particular, we investigate whether the covalent bounding is constrained by the surface energy, surface chemistry or morphology. To achieve this, we will investigate the correlation of covalent binding with the presence of oxygen-containing groups, surface roughness and surface energy determined with contact angle.

We apply PIII modification to low density polyethylene (LDPE) and ultra high molecular weight polyethylene (UHMWPE), and attach HRP [55] a widely used enzyme with available biological, crystallographic and spectroscopic data. The polyethylene polymers are used in applications such as medical devices, including human body prosthetics, where protein attachment plays important role. Polyethylene when untreated is a representative hydrocarbon polymer with a high hydrophobicity. The modification of polyethylene by ion beam implantation [56-72] has been investigated.

## 2. Experiment

Polyethylene (PE) film of low density (LDPE) and ultrahigh molecular weight (UHMWPE) was purchased from Goodfellow. PE strips of 2x5 cm$^2$ area and 0.200, 0.020 and 0.075 mm thickness were washed with ethanol and dried minimum one day before using.

Horseradish Peroxidase (HRP) cat.No.P6782 and all other chemicals were purchased from Sigma. The nitrogen gas used for PIII was 99.99% pure.

Plasma immersion ion implantation was carried out in an inductively coupled radio-frequency plasma powered at 13.56 MHz. The plasma power was 100 W with reverse power of 12 W when matched. The plasma density during treatment was monitored by a Langmuir probe with rf-block from Hiden Analytical Ltd. The base pressure was $10^{-6}$ Torr (~$10^{-4}$ Pa). The pressure of nitrogen during implantation was $2 \cdot 10^{-3}$ Torr ($4.4 \cdot 10^{-2}$ Pa) and the pressure of argon during implantation was $3.3 \cdot 10^{-4}$ Torr ($7.3 \cdot 10^{-3}$ Pa). This difference of pressure provided the same plasma density for nitrogen and argon plasma. Acceleration of ions from the plasma was achieved by the application of high voltage (20 kV) bias pulses of 20 μs duration to the sample holder at a frequency of 50 Hz. The thickness of modified layer of polyethylene is 70 nm, as it was calculated in [37, 38].

The samples were mounted on a stainless steel holder, with a stainless steel mesh, electrically connected to the holder of 150 mm diameter, placed 45 mm in front of the sample surface. In experiments with chemical vacuum chamber, the sample holder of 25 mm diameter without mesh was used.

The ion fluence was calculated from the number of high voltage pulses multiplied by the fluence corresponding to one pulse. The fluence of one high voltage pulse was determined by comparing UV transmission spectra from polyethylene films implanted under conditions used here to samples implanted with known fluences in previous PIII and ion beam treatment experiments [38].

The fluence rate for wide sample holder with mesh was $10^{13}$ ions/cm$^2$ per second and for small sample holder without mesh was $4 \cdot 10^{13}$ ions/cm$^2$ per second of PIII treatment time. The samples were treated for durations of 20 - 800 secs, corresponding to implantation ion fluences of $(0.5 – 10) \cdot 10^{15}$ ions/cm$^2$.

For chemical post-treatment without contact with air, the chemical vacuum chamber connected to the plasma chamber was used. Chemical chamber was closed from plasma chamber during plasma treatment (Fig.1). Protein solution or acrylic acid solution in metal bath was placed into chemical chamber, pumped up to vapour pressure of the solution (20 Torr). Then the chamber was vented by nitrogen up to atmospheric pressure (750 Torr). This cycle of pumping and venting of chemical chamber was repeated 5 times to remove the atmospheric oxygen from the chemical chamber up to $2.4 \cdot 10^{-7}$ % or $2 \cdot 10^{-6}$ Torr (estimation). Then the bath with solution was removed into bottom part and closed from main volume of the chemical chamber. Then the chemical chamber was pumped up to pressure of plasma chamber.

The sample was moulded on metal holder with diameter of 25 mm, placed into plasma chamber and treated. After PIII treatment the sample was taken by mechanical arm and moved into chemical chamber at residual pressure of $10^{-4}$-$10^{-5}$ Torr. Then the plasma chamber was closed from chemical chamber. The chemical chamber was filled by nitrogen up to pressure of 20 Torr and the bath with solution was opened and moved up to the chemical chamber. The sample was immersed into the bath with solution. After exposure in solution the chamber was opened and the sample was used for an analysis.

The wettability of PE was measured using the sessile drop method. Kruss contact angle equipment DS10 was employed to measure the contact angles. De-ionised water and Diiodomethane were dropped on the sample and the angle between edge of drop and the surface was measured. Surface energy and its components (polar and dispersic parts) were calculated using the Rabel model with regression method.

Optical microphotographs of the polyethylene surface were made with microscope Axioskop, Carl Zeiss Jena attached with CCD camera KS-100-3.0. Zeiss software was used for image treatment. The 3D-images of sample surface were obtained using an atomic force microscope (AFM) Rasterscope C-21 (DME, Denmark) with software Dual Scope/Rasterscope SPM 1.3.2. The measurements were done in FZR, Germany.

After PIII treatment, the PE samples were incubated in HRP solution (50 μg/ml in 10mM sodium phosphate buffer pH 7) overnight at 23ºC. In kinetics study, the time of incubation varied. After incubation, PE samples were washed six times (20 minutes each wash) in buffer (10 mM sodium phosphate buffer pH 7). Samples for FTIR ATR spectral analysis were washed in de-ionised water for 10 seconds to remove buffer salts from the PE surface.

HRP activity was measured by two methods. In first method, the PE sample (13 mm x 15 mm) was clamped between two stainless steel plates separated by an O-ring (inner diameter 8 mm, outer diameter 11 mm) which sealed to the plasma treated surface. The top plate contained a 5 mm diameter hole. TMB assay was added to the polymer surface. After 30 seconds 25 μl was removed and added to 50 μl of 2M HCl followed by 25 μl of unreacted TMB to make the volume up to 100 μl. Optical density was then measured at a wavelength of 450 nm using a DU 530 Beckman spectrophotometer.

In second method, the PE quadrant sample of 1 cm$^2$ was placed on bottom of UV spectrophotometer cuvette. The micromixer with 10 mm length and 0.5 mm diameter mixer bar and elec-trical micromotor was placed on the top of the cuvette. Speed rate of mixer was 100 rpm. The cuvette was placed in spectrophotometer and TMB solution was added to the cuvette. The spectrophotometer recorded optical density at 635 nm wavelength in kinetics mode.

The protein solution of 1 ng/ml concentration was prepared by dilution of concentrated solution for analysis of catalytic activity of protein corresponding to monolayer of protein. Such low concentration can give error because of protein attachment on walls of the glass and polymer volumes used for dilution. The BSA sealing of the surfaces of volumes at dilution was used against the protein attachment.

FTIR-ATR spectra from the PE samples were recorded using a Digilab FTS7000 FTIR spectrometer fitted with an ATR accessory (Harrick, USA) with trapezium Germanium crystal and incidence angle of 45º. To obtain sufficient signal/noise ratio and resolution of spectral bands we used 500 scans and a resolution of 1 cm$^{-1}$. Differences, obtained by subtraction, between spectra of samples before and after PIII treatment were used to detect changes associated with the surface treatment. All spectral analysis was done using GRAMS software.

FTIR ATR spectra of polyethylene were recorded in-situ after PIII treatment. The plasma chamber was connected with spectral chemical chamber (Fig.1). The infrared beam from spectrometer passed through KRS-6 window into spectral chamber with ATR accessory. After ATR accessory, the beam returns to spectrometer through the KRS-6 window to detector. The spectral chamber was under the same vacuum during measurements as the plasma chamber. The polyethylene sample was placed on high voltage electrode in plasma chamber and treated there. The treated polyethylene sample was replaced by mechanical arm from high voltage electrode of plasma chamber to ATR crystal in spectral chamber without breaking of vacuum. Therefore, the sample was not exposed by air between PIII treatment and FTIR ATR spectra recording. The ATR crystal of 2 mm diameter had hemispherical geometry. Because of focused infrared beam the angle of beam incident was different and the conditions of attenuated total reflection were not fulfilled for all beam angles. It provided high sensitivity of the analysis, but quantitative measurements were not possible. For comparison, the same ATR attachment was used at presence of air in the spectral chamber and the treated polyethylene sample was exposed on air after PIII treatment.

Micro-Raman spectra were recorded in a backscattering mode excited by Nd:YAG laser irradiation (2 ω, wavelength = 532.14 nm) on a diffraction double-monochromator spectrometer (HR800, Jobin Yvon) with the LabRam 010 system (Lastek, Adelaide, SA, Australia). An optical microscope was used for focusing the exciting laser beam and for collecting the Raman scattered light. An objective of 100 units was used. The sample surface position alignment during the Raman signal acquisition was adjusted according to an image projected onto a screen from a microscope-mounted charged coupling device camera. The intensity of the laser beam was adjusted to avoid overheating of the samples. A spectral resolution of 4 cm$^{-1}$ was used. The number of scans was selected from the range of 100–4000 to provide a sufficient signal/noise ratio for the spectra. LabRam software was used for spectral analysis.

XPS measurements were done with a spectrometer (Specs, Berlin, Germany) equipped with an Al X-ray source with a monochromator operated at 200 W, a hemispherical analyzer, and a line delay detector with nine channels. Survey spectra were acquired for binding energies in the range of 0–1200 eV at a pass energy of 30 eV. $C_{1s}$, $O_{1s}$, and $N_{1s}$ region spectra were acquired at a pass energy of 23 eV with 10 scans to obtain a higher spectral resolution and to lower the noise level. $S_{2p}$ region spectra were acquired at a pass energy of 30 eV with 50 scans and a 0.5-s dwelling time to obtain even lower noise levels. CASA software was used for spectral analysis.

## 3. Results and discussion

### 3.1. Structure changes in polyethylene surface layer resulting from PIII treatment

The effect of ion bombardment during PIII was readily observed as a darkening of the treated polyethylene. The dark colour corresponds to carbonization of the surface layer of polyethylene and increases with ion fluence. The dark colour remains after washing of the modified polyethylene surface with water, buffer and the organic solvents toluene, ethanol and acetone. The carbonization was studied by Raman and UV-vis spectroscopy.

The micro-Raman spectrum of the modified surface layer shows peaks which are interpreted as vibrational modes of pure carbon structures and don't show any lines attributed to vibrations of polyethylene macromolecules (Fig.2). The spot of the focused laser beam in the Raman spectrometer is about 10 μm, larger than the thickness of the modified layer of polyethylene (70 nm). The intensity of the carbon lines is significantly higher than the intensity of the polyethylene lines because of a resonance Raman effect which is observed because the laser wavelength is resonantly absorbed by the electrons in the carbon structures. The spectrum is fitted by individual Gauss functions with maxima at 1587 $cm^{-1}$ and 1350 $cm^{-1}$. These lines are interpreted as $E_{2g}$ vibrations ("G" peak at 1587 $cm^{-1}$) and $A_{1g}$ vibrations ("D" peak at 1350 $cm^{-1}$) of graphitic structures [73, 74]. Using known correlations of the ratio of Raman peak intensities I(D)/I(G) and the D and G peak positions with the characteristic size of graphitic structural domains [75], the carbonised layer is found to consist of graphitic islands of a characteristic size of 1 nm.

Quantitative analysis of the extent of carbonization can be extracted from the UV-Vis spectra of the polyethylene film. Untreated polyethylene has an absorption at short wavelengths of less than 250 nm. The UV-vis transmission spectra of PIII treated polyethylene films shows a broad absorption (Fig.3) with a shoulder that progressively shifts to the red and increases in intensity as the fluence of ions increases.

These changes of absorbance can be viewed from two different standpoints. The absorbance is caused by unsaturated carbon structures including condensed aromatic rings and polyene groups. An increase in size of the conjugated regions of unsaturated carbon structures causes a red shift in their absorption Therefore, the observed red shift and increase of intensity of the absorption shows that there is a progressive increase of the unsaturated group concentration and an in the size of conjugated structures with increasing fluence of PIII treatment. For example, the shoulder of absorbance for polyethylene after 1600 sec of PIII treatment is observed at 2.5 μm wavelength, corresponding to large islands consisting of conjugated aromatic rings arranged in graphite-like structures – nanographite islands. Using the Woodward-Fieser rule [76, 77], the absorbance at 2.5 nm wavelength corresponds to graphitic island containing about 40 conjugated aromatic rings, that would have characteristic size of about 1.5 nm. This value is comparable with the 1 nm size of graphite islands estimated from the Raman spectra.

The shift of the absorbance corresponds to decrease of the gap between energy bands corresponding to valence electrons and conduction electrons. The well known Tauc formula can be used to calculate the energy gap in ion beam implanted polymers, as shown in reference [38].

With increasing PIII fluence the energy gap narrows from 3 eV for untreated polyethylene to 0.6 eV for polyethylene treated for 1600 sec.

The appearance of unpaired electrons associated with free radicals in modified polyethylene is observed in electron spin resonance (ESR) spectra as shown in (Fig.4). The untreated polyethylene film contains a low concentration of unpaired electrons. Free radicals in as received polymer can be generated by environmental factors such as light, temperature and oxygen species (ozone, peroxide and excited oxygen molecules) as well as trapped residual free radicals remaining after the polymerisation process. The ESR intensity of the untreated sample shown in Fig.4 corresponds to the natural concentration of free radicals in the bulk sample of 50 μm thickness.

The principal contribution to the ESR spectrum of the treated sample is attributed to the thin modified surface layer of 70 nm thickness. Allowing for the difference in thickness of the polyethylene film (50 μm) and the modified layer (70 nm), the observed increase in the ESR signal for the PIII treated sample of 160 times corresponds to a factor of $10^5$ times greater free radical concentration in the PIII treated surface layer compared to the untreated material. The intensity of the ESR spectrum decreases slowly with storage time of the sample at room temperature, showing that the unpaired electrons recombine as a result of free radical reactions.

The g-factor (0.0028) of ESR signal of PIII treated polyethylene is close to the g-factor of free electrons (0.0025), and corresponds to the g-factor unpaired free electrons delocalised on graphite structures [78-82]. Original position of such electrons can be on edges of graphite structures. Usually, the unpaired electrons have short life time under room temperature conditions due to high activity. In the case of PIII treated polyethylene, the saving of the free radicals is due to delocalisation of unpaired electron on π-electrons of graphite structure so, that free radical does not have a possibility to be terminated by hydrogen, oxygen or other active particles placed near the graphite islands. The aromatic ring π-electrons stabilise these unpaired electrons.

The FTIR ATR spectra of treated polyethylene show significant changes in the surface layer after PIII treatment (Fig.5). The broad continuous band from 1700 to 700 cm$^{-1}$, interpreted as vibrations in disordered carbonised layers, appears after PIII treatment. The intensity of this band grows with the duration of the PIII treatment. Vibrational modes of this carbonised layer have very broad frequency distribution. The vibrational states corresponding to individual carbon structures are not resolved.

Some narrow vibration lines related to new structures are observed in the spectra of treated samples. These lines in the region of 1750-1600 cm$^{-1}$ correspond to ν(C=O) vibrations in carbonyl, carboxyl, aldehyde, lactone and ester groups, and ν(C=C) vibrations in unsaturated hydrocarbons fragments and aromatic rings. Narrow peaks at 921, 907 and 887 cm$^{-1}$ corresponding to vinyl, vinylidene and vinylene groups respectively are observed. The broad line in the region 3600-3200 cm$^{-1}$ corresponds to ν(OH) vibrations of hydroxyl, carboxyl and peroxide groups.

All these spectral changes caused by the PIII treatment are observed as additional features superimposed on the intense vibrational spectra of unmodified polyethylene macromolecules, which consist of asymmetrical and symmetrical stretch vibrations ν(C-H) at 2917 and 2860 cm$^{-1}$, bending vibrations in –CH$_2$- groups δ(C-H) at 1462 cm$^{-1}$ and in –CH$_3$ branch end groups at 1375 cm$^{-1}$, and out-of-plane vibrations γ(C-H) at 720/730 cm$^{-1}$ doublet. The intensity of the lines related to unmodified polyethylene macromolecules is higher than the intensity of lines of the newly introduced groups. The difference in intensity corresponds to the ratio of the modified layer thickness (70 nm) and to the depth of the FTIR ATR analysis (700-400 nm, depending on wavenumber).

The appearance of lines corresponding to oxygen containing groups (C=O and OH) shows that the oxidation of the PIII modified layer occurs after exposure of the treated polyethylene samples to air. Residual oxygen in the plasma chamber is expected to have a partial pressure of not more than $10^{-5}$ torr. Therefore, the oxygen incorporation into modified polyethylene structure could be expected to occur at a rate $10^{-7}$ time less in plasma chamber than in air.

Fig. 6 shows the ATR-IR spectrum of the non-oxidized PIII treated surface observed in-situ before contact of the sample with air. These spectra show that the modified layer of polyethylene does not contain any oxygen containing groups When the sample is exposed to air after treatment, the spectra of the same sample shows the $\nu$(C=O) and $\nu$(OH) lines of oxygen containing groups. We conclude that oxygen containing groups appear in the surface layer as result of reaction with air, not with residual oxygen in the plasma chamber.

The intensity of the oxygen containing group lines increases with time after modification. For example, the dependence of the absorption due to carbonyl group vibrations at 1712 cm$^{-1}$ on storage time after PIII is presented in Fig.7. The normalized absorbance of the $\nu$(C=O) line increases with time of storage in air. This means that the concentration of carbonyl groups in the surface layer grows with time. The saturation level of the carbonyl group concentration depends on the ion fluence (PIII time). The maximal concentration is observed for median times of PIII treatment (80 sec), corresponding to an ion fluence of $10^{15}$ ions/cm$^2$. The same effect was observed for PIII treated polystyrene when oxidation was maximal at such an ion fluence, which provides the highest level of macromolecule destruction but not a fully carbonised surface layer [43]. This means, that oxidation primarily results from the interaction of free radicals created by damage to the macromolecular chains with atmospheric oxygen. The appearance of carbonised structure decreases the level of oxidation due to stabilization of free radicals on $\pi$-electron clouds as discussed above.

The oxidation process is facilitated by on exposure to oxygen but not on exposure to water. The storage kinetics of PIII treated polyethylene stored under water show the same concentration curve of oxygen-containing groups as for modified polyethylene sample stored in air. Storage under water does not cause any increase in OH For example, the dependence of $\nu$(OH) line absorbance at 3400 cm$^{-1}$ wavenumber is presented in Fig.8. The samples were treated by PIII and immediately placed in μQ-water. The air exposure time between venting of the PIII chamber by air and placing in water was 2-3 minutes. This is a relatively short time in comparison with the kinetics of the process of some days. The oxidation under water occurred due to the oxygen dissolved in water. The samples were kept under water before spectrum measurement. Removed from water sample was kept on air and measured by spectra. After complete evaporation of water from the surface layer of polyethylene (up to 3 hours), the $\nu$(OH) absorbance points lay on the same curve as for oxidation kinetics on air. The position of maximum and shape of the line is similar for the samples stored in air and under water. This means that storage under water does not change the oxidation kinetics and that the presence of water during oxidation does not play a role. The same similarity of the oxidation kinetics is observed using $\nu$(C=O) vibration line.

The changes of chemical structure have been observed by XPS spectra. The spectra of untreated polyethylene has singlet $C_{1s}$ line and low intensive $O_{1s}$ line (0.1 % from carbon line intensity) showing the oxidation of polyethylene surface under environmental factors such as light, atmospheric oxygen and cosmic rays (Fig.9). After PIII treatment, the spectrum shows $C_{1s}$, $O_{1s}$ and $N_{1s}$ multiplet lines. The fitting of these lines gives separate peaks of new bonds such as C-O, C=O, C-N, C=N, N-O and N=O. The appearance of oxygen (Fig.9a) in high concentration (10 % of total elements) is consistent with the new oxygen containing groups observed by FTIR spectra of PIII treated polyethylene and with previous results [43]. The nitrogen in polyethylene after PIII with nitrogen ions can be explained with capturing of implanted nitrogen ions in modified layer. The nitrogen (2 % of total elements) in polyethylene after PIII with argon ions (Fig.9b) can be only explained by reactions of modified polyethylene with atmospheric nitrogen after exposure of the treated samples in air.

The structure transformations in polyethylene surface layer during and after PIII can be explained by free radical reactions. There is extensive literature discussing free radical reactions in polyethylene resulting from different kinds of radiations including ion beam implantation, UV, VUV, electron beam, gamma-irradiation, X-ray irradiation [83-85]. Examples of reaction induced by ions (i) are shown in (1). Many of these reactions involve free radicals (R•) created by ions.

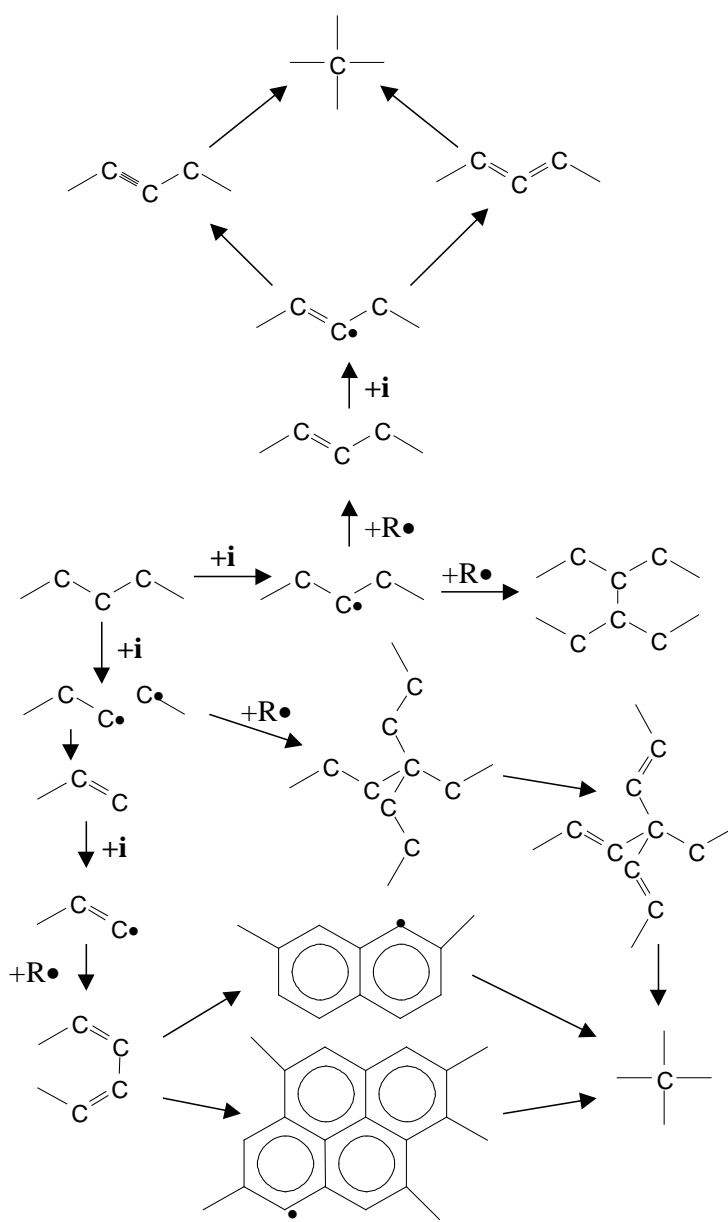

(1)

The scheme shows, that initial polyethylene structure becomes enriched in double bond groups, condensed aromatic rings, crosslinks and amorphous carbon. At very high fluences the structure is transformed into highly crosslinked amorphous carbon.

After exposure of the treated surface to the atmosphere, oxidation reactions may follow, such those shown in (2):

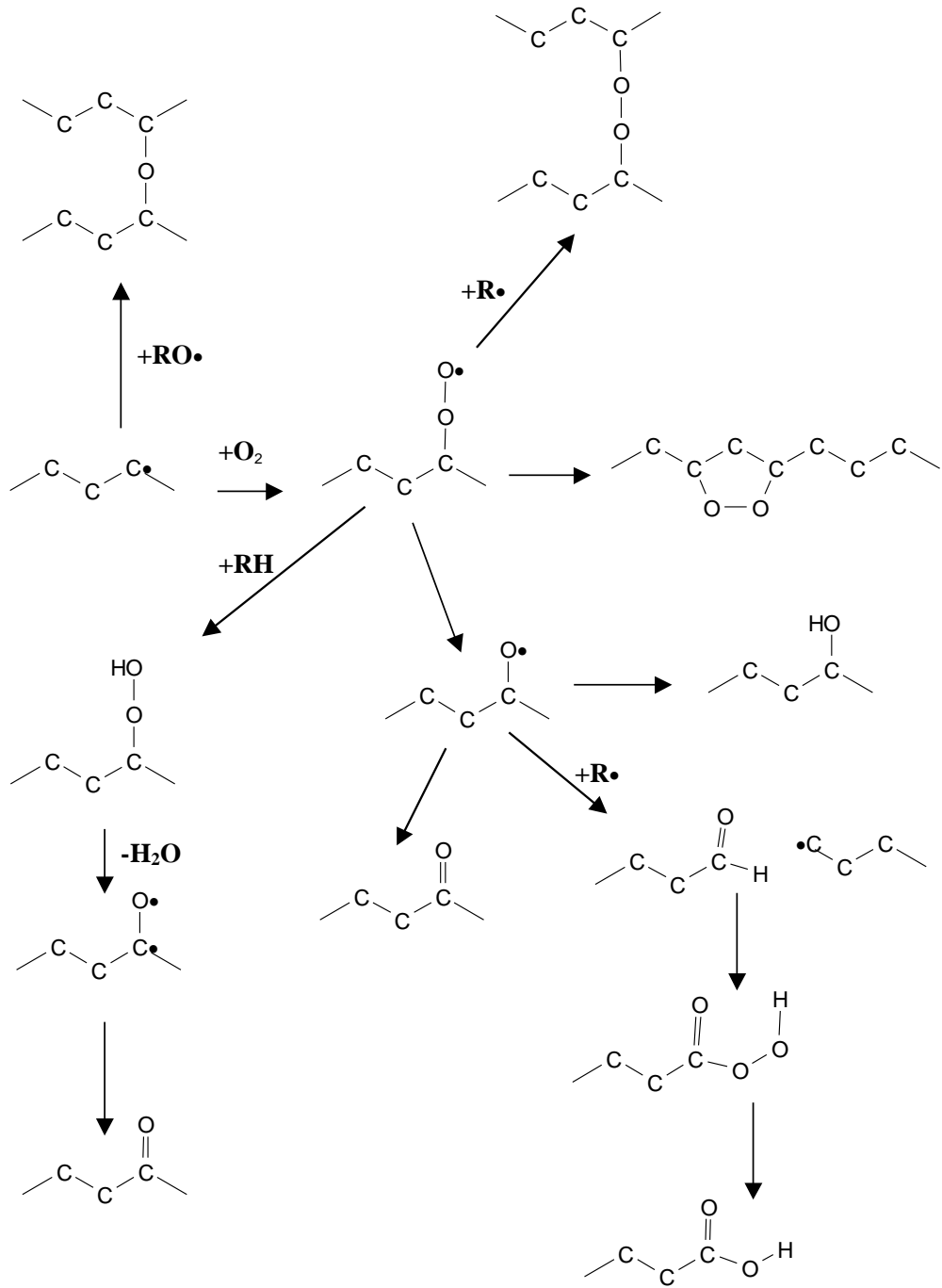

(2)

The stable oxygen-containing groups such as carbonyl, carboxyl, aldehyde, hydroxyl, ester, ether are created.

After exposure of the treated surface to the atmosphere, the reactions of atmospheric nitrogen with multiradicals may follow, such those shown in (3):

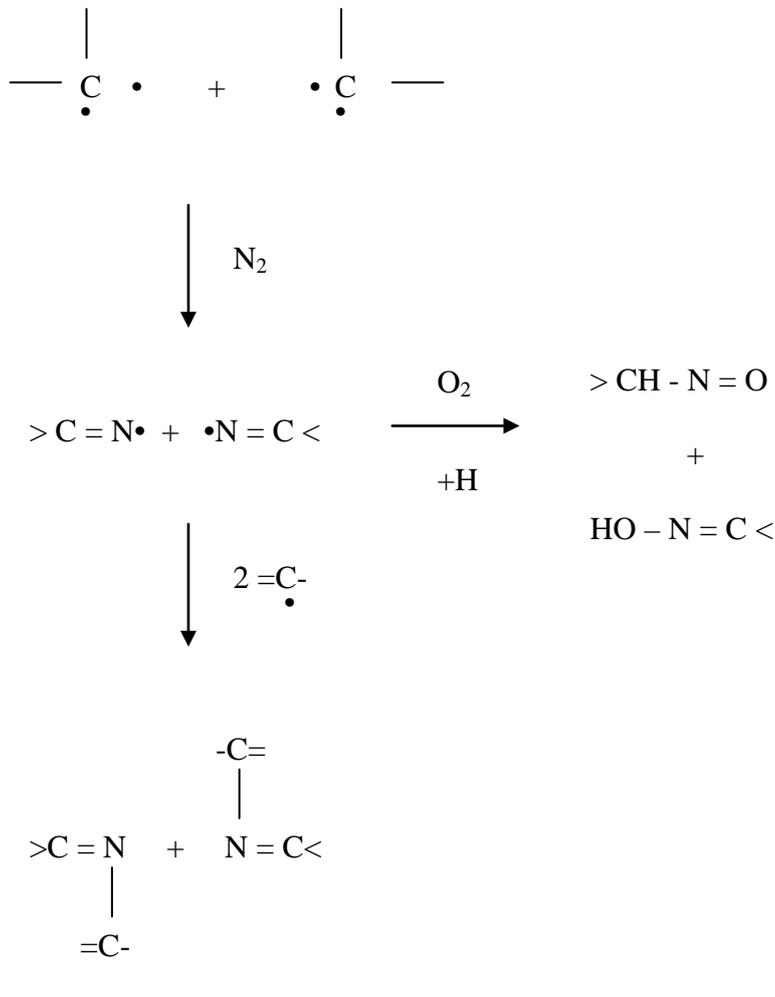

(3)

Carbon structures containing $sp^2$ and $sp^3$ hybridisations, oxygen-containing groups and unsaturated carbon-carbon bonds in macromolecules are observed in FTIR and Raman spectra as shown above. Some free radicals remain active in the modified surface layer even for long storage times after the PIII treatment. The chemical structure of the modified polyethylene surface explains its reactivity in many different kind of chemical reactions.

The surface of polyethylene after PIII becomes textured with corrugations as observed using optical microscopy and using AFM in scanning mode (Fig.10). The lateral wave period is 2.3±0.2 μm and the peak to peak amplitude of the wave is about 0.35±0.05 μm. Assuming the corrugations are sinusoidal, the effective surface area of the polyethylene increases by 1.2±0.1 times after PIII. The reason for the corrugations structure may be internal stresses caused by carbonisation of the surface layer, dehydration processes and crosslinking. All of these processes decrease the volume of the modified layer. The direction of the corrugations may be influenced by residual stresses generated during the rolling of the polyethylene films. In our experiment, the modified layer is 70 nm, much less than the unmodified PE layer, so that relaxation of the internal stresses is accommodated in the modified layer.

## 3.2. Surface energy of PIII modified polyethylene

The untreated polyethylene surface is hydrophobic with a wetting angle of a water drop on untreated polyethylene surface of higher than $90^0$ degrees. Immediately after PIII treatment, the water contact angle becomes very low, near $24^0$ degrees i.e. very hydrophilic (Fig.11). The same sharp changes of wetting angle are observed for formamide, methyleneiodide and glycerol drops. The total surface energy increases from 24 to 67 MJ/m$^2$ (Fig.12). The contributions of the polar and dispersive parts of the total surface energy are different. The polar part of the surface energy increases significantly from 4 MJ/m$^2$ to 44 MJ/m$^2$. The dispersive part does not change much from 20 MJ/m$^2$ to 22 MJ/m$^2$.

The wetting angle and the surface energy of PIII modified polyethylene are not stable with time of storing after treatment. The water contact angle increases to 60-70$^0$ degrees after 2 weeks of storing time on air. The total energy decreases to 44 MJ/m$^2$, the polar part of the surface energy decreases to 14 MJ/m$^2$. The dispersive part of the surface energy increases slightly to 30 MJ/m$^2$.

The increase in surface energy is caused by chemical transformations in the modified polyethylene surface, in particular, the appearance of high energy polar groups. The same increase is observed after plasma treatment, UV light, electron and γ-irradiation [10, 14, 46]. The decrease in surface energy observed after storage of the modified material may have the same origin as similar surface energy decreases observed after plasma polymerisation or plasma treatment of polymer surfaces [86, 87].

The surface energy decrease and recovery of hydrophobicity with time in the plasma treated polymers has been attributed to the reptation or diffusion of the surface polymer chains and the rotation of high energy surface functional groups back into the polymer bulk [47]. Surface energy is thereby reduced and hydrophobicity increased. However, for a PIII treated surface layer the diffusion of polymer chains as in the reptation picture should be hindered by the dense crosslinking associated with highly carbonised structures appearing after high fluence modification.

The explanation of rotation of high energy surface functional groups at the polymer surface in polar media like water is not observed in our case of PIII treated polyethylene. Although the contact angle reduce from 60-70$^0$ degrees to 51$^0$ by immersion in water, our FTIR data (Fig.8) shows that this is associated by uptake of water rather than the rotation of high energy groups. The concentration of –OH groups clearly increases upon contact with water and then gradually decreases again. The water molecules remain in PIII modified layer longer time than required to evaporate drop of water from the surface. This is possibly connected with absorption of water molecules on oxygen containing groups and free radicals, which would cause a slow diffusion of the water molecules from the modified surface layer. After complete removing the water molecules as evidence by the reduction of the OH line intensity, the water contact angle increases again to 56$^0$. Therefore, the recovery of high wettability under exposure in water can be explained by diffused water in the modified layer and the low contact angle after immersion is due to strong contribution of water-water interactions in the surface layer.

Another possible explanation for hydrophobic recovery may be via the adsorption of adventitious carbon and hydrocarbons on the activated surface during atmospheric exposure [48]. The changes in wettability in this study were repeatable despite the variable nature of the uncontrolled laboratory atmosphere and the FTIR-ATR spectra did not show any significant growth in lines attributed to adsorbed hydrocarbons on the surface. Besides that, the decay of the water wetting angle is the same in air and in vacuum under $10^{-5}$ Torr pressure. In the last case, an adsorption of adventitious carbon and hydrocarbons on the activated surface is excluded or decreased. Therefore, the adsorption mechanism of the decay can be excluded.

We believe that the changes in surface energy observed with storage time after modification are associated with chemical changes in the surface of the modified polyethylene. The character and kinetics of these transformations is complex and depend on a number of reactions characterized by different rates of reaction. The macrokinetics may be applied to simulate these transformations. These relaxation processes may be fitted using exponential functions that correspond to first order reactions. Exponential functions were fitted to the total and polar component of the surface energy and agreement between experimental data and the theoretical curve was found (Fig.12). The fitted curve is a biexponential function of the form:

$$\sigma = \sigma_1 \cdot \exp(-\frac{t}{t_1}) + \sigma_2 \cdot \exp\left(-\frac{t}{t_2}\right) + \sigma_\infty \qquad (1)$$

where $t_1$ and $t_2$ are the characteristic times of the decay, $\sigma_\infty$ is the surface energy of the modified polyethylene after infinite storage time and the $\sum(\sigma_i+\sigma_\infty)$ value is the surface energy of modified polyethylene immediately after PIII treatment. After fitting of experimental points, the equations for polar and dispersive parts of the surface energy are given by the following:

$$\sigma_{dispersive} = 7.5 \cdot \exp(-\frac{t}{5000}) - 18 \cdot \exp\left(-\frac{t}{60}\right) + 31 \qquad (MJ/m^2) \qquad (2)$$

$$\sigma_{dispersive} = 15 \cdot \exp(-\frac{t}{60000}) + 27 \cdot \exp\left(-\frac{t}{50}\right) + 4 \qquad (MJ/m^2) \qquad (3)$$

where t is in minutes. The characteristic time shows that the wettability of the modified polyethylene is stabilized 3-4 days after PIII treatment. The total surface energy immediately following PIII treatment is 67 $MJ/m^2$ with polar component of 46 $MJ/m^2$. The surface energy after infinite storage time (35 $MJ/m^2$ with a 4 $MJ/m^2$ polar component) is not the same as the value for the untreated polyethylene surface (23.8 $MJ/m^2$ with a 3.7 $MJ/m^2$ polar component). The change in surface energy is determined mostly by the polar component. The dispersive component of the surface energy remains in the range of 20-30 $MJ/m^2$ for untreated and for modified polyethylene surfaces.

The total surface energy immediately following PIII treatment (67 $MJ/m^2$) is much higher than normally reported for an equilibrium property of a polymer which show total surface energies not exceeding 45-50 $MJ/m^2$ [88]. There are two possible sources for the high surface energy: newly introduced oxygen containing polar groups and free radicals. An examination of the literature shows that surface energies as high as 67 $MJ/m^2$ cannot be explained by a coverage of the new polar groups that we observe. The surface energies reported for carbonyl containing group all have surface energy below 50 $MJ/m^2$ [88] e.g. polymethylmethacrylate with high concentration of C=O groups in the backbone (49 $MJ/m^2$), polycarbonate with aromatic rings and carbonyl groups in the backbone (46.7 $MJ/m^2$) and polyetherester ketone PEEK with ether, ester and aromatic ring groups in the backbone (46 $MJ/m^2$). In the case of hydrogenated and oxidized surfaces of diamond and graphite, surface energies of 47 MJ/m$^2$ and 65.8 MJ/m$^2$ respectively have been reported [89]. A low surface energy is observed for fullerenes which do not have unpaired electrons at the surfaces (48.1 MJ/m$^2$) [90] and carbon nitrides (51.9-41.9 MJ/m$^2$) [91].

The surface energy of carbon structures like graphite and diamond is sensitive to a presence of dangling bonds at the surfaces. The surface energy of carbon structures terminated with unpaired electrons is reported to be in range of 4000 MJ/m$^2$ to 1000 MJ/m$^2$ [92, 93]. This high surface energy arises from the presence of free radicals in the surface layer. We do not expect such a high coverage of the surface of our modified polyethylene by free electrons associated with free radical groups, however, the high surface energy does indicate that the PIII treatment results in the generation of a significant number of free radicals at the freshly treated polymer surface. The presence and disappearance of free radicals on the modified polyethylene surfaces changes the polar component of the surface energy, not the dispersive component, as expected of the strong polar electrostatic interactions of free uncoupled electrons.

Therefore, we propose that most of the changes of wettability and surface energy of the PIII-modified polyethylene are produced by free radicals in the top layer that is in contact with the wetting liquid. Residual differences in surface energy for untreated and PIII modified polyethylene after long storage times are connected with stable groups that appear after the completion of the free radical reactions.

The presented above results show, that the PIII treated surface has been changed dramatically. The surface becomes richer with polar intermolecular interactions; the chemical groups such as hydroxyl, carbonyl, carboxyl, aldehyde, ester, ether, carbon unsaturated groups, aromatic condensed structures and the number of free radicals change chemical activity of the surface layer; changes in morphology of the surface gives higher effective surface area.

### 3.3. Improved attachment of protein on PIII treated surfaces

The untreated polyethylene surface during incubation in buffer solution with HRP attaches protein molecules. The FTIR ATR spectrum of an incubated polyethylene surface after subtraction of the spectrum of polyethylene incubated in buffer solution without protein shows the presence of characteristic lines 3300 cm$^{-1}$ of Amide A, 1650 cm$^{-1}$ of Amide I and 1540 cm$^{-1}$ of Amide II related to strongly absorbing vibrational modes of the protein molecules (Fig.13). The spectrum of PIII modified polyethylene incubated in protein solution shows the same lines. However, the protein lines in the spectra of the PIII treated surface are more intense than in the spectra of untreated polyethylene sample. We have developed a normalisation procedure for a quantitative measure of the amount of attached protein using all three Amide lines.

The intensity of the protein lines is significantly lower than that of the vibrational lines of the underlying polyethylene. This is consistent with the protein layer being thinner than the penetration depth of the infrared beam. Therefore, the intensity of polyethylene lines can be used as an internal normalization standard for the protein spectrum. According to Bouger-Lambert-Beer law, an absorbance (A) for infrared transmission spectra is given

$$A = \varepsilon \cdot C \cdot h \qquad (4)$$

where C is concentration of the absorbing groups, h is thickness of the layer and $\varepsilon$ is extinction coefficient, the structure effect can be analyzed by changes of the concentration C of the groups. In the case of ATR spectra, the depth penetration of the infrared beam into a sample immediately adjacent to the crystal is

$$h = \frac{\lambda}{\sqrt{n_{21}^2 - \cos^2(\theta)}} \quad (5)$$

where h is the depth penetration of the infrared beam into the sample, $n_{21}$ is a ratio of refractive indexes of ATR crystal and the sample, Lambda is the wavelength of infrared beam. An effective amount of attached protein P can be calculated using

$$P = \frac{\sum_i k_i \cdot \frac{A(Amide_i)}{A(CH_2)}}{3} \quad (6)$$

where $A(Amide_i)$ is the absorbance of i Amide line, $A(CH_2)$ is the absorbance of -$CH_2$- stretch vibration line at 2917 cm$^{-1}$ or the bending vibration line at 1462 cm$^{-1}$, $k_i$ is normalising coefficient as ratio of Amide-i/Amide I extinction coefficients multiplied by the ratio of penetration depth of the infrared beam at the Amide-i wavenumber and the Amide I wavenumber. The amount of attached protein on PIII modified polyethylene is approximately doubled in comparison with untreated polyethylene. The same difference was observed in our previous experiments with polystyrene [41], Teflon [40] and Nylon [42].

The presence of protein is also observed by XPS spectra. XPS spectra of PIII treated polyethylene surfaces with and without attached HRP protein are presented in Figs.14(a-d). The samples after incubation in protein solution were rinsed with μQ-water, dried and placed in the vacuum chamber of the XPS spectrometer. The XPS spectra were recorded in $C_{1s}$, $O_{1s}$, $N_{1s}$ and $S_{2p}$ regions. The HRP protein monolayer is not expected to be higher than 7-9 nm, therefore the X-ray beam will penetrate through the protein layer and enter the polymer surface even when the polymer is completely covered by a protein layer. Therefore, the spectra show the protein layer peaks as well as the polymer surface peaks in different proportions. Subtraction of the XPS spectra of the polymer surface will give spectra of the protein layer only.

The $C_{1s}$ carbon peak of the PIII treated polyethylene surface (Fig.14a) is a multiplet due to the presence of oxygen-containing and nitrogen-containing groups. The $C_{1s}$ peak at 285 eV has a long tail on the high energy side of the main peak. The spectrum of PIII treated polyethylene with attached HRP shows clearly visible shoulders at 286.7 and 288.3 eV. The fitting analysis of this tail is problematic, because of high variability of the different oxygen-containing and nitrogen-containing groups in the PIII treated surface layer. However, the subtraction procedure can give separated protein spectrum with 286.7 and 288.3 eV peaks attributed to C-N and C=O groups accordingly. The same peaks are observed in XPS spectra of HRP prepared as a thick layer on a silicon wafer substrate.

The XPS spectra of untreated polyethylene do not show a nitrogen ($N_{1s}$) line while PIII treated polyethylene contain a small amount of nitrogen and shows a nitrogen peak (Fig.14b). The presence of such a nitrogen peak could be explained by reaction of free radicals with implanted nitrogen and nitrogen absorbed from the air after PIII treatment. The attachment of HRP protein gives an additional peak at 400.3 eV on the top of the $N_{1s}$ peak of PIII treated polyethylene. The subtraction reveals the protein $N_{1s}$ peak as shown in Fig.14b.

The oxygen $O_{1s}$ peak appears in XPS spectra of polyethylene only after PIII treatment as the result of oxidation of the surface layer in air. The peak consists of two individual peaks for oxygen in C-O and C=O bonds (Fig.14c). The presence of HRP protein on the surface increases the intensity of the $O_{1s}$ peak as HRP contains oxygen. The subtracted spectra also show peaks at 531.85 and 533.05 eV attributed to oxygen in C-O and C=O bonds of HRP respectively.

The presence of sulphur-containing aminoacids Cysteine and Methylcysteine in HRP protein gives rise to the $S_{2p}$ peak in the XPS spectra of PIII treated polyethylene with attached HRP (Fig.14d). The HRP protein contains 10 Cysteine and 7 Methylcysteine residues that contribute 17 sulphur atoms to one HRP molecule. The structure of the HRP molecule allows the amount of sulphur to be calculated at 0.54 atomic %. The XPS spectrum of HRP molecule on PIII treated polyethylene surface shows 0.16 atomic % of sulphur calculated from the relative ratio of the $S_{2p}$ peak intensity to the sum of $C_{1s}$, $O_{1s}$ and $N_{1s}$ peak intensities. This is consistent result, because the X-ray beam of XPS spectrometer penetrates deeper into the sample than the maximum available thickness of HRP monolayer (the longest dimension of the HRP molecule is 7nm). The XPS spectra of the sample without HRP attachment does not show $S_{2p}$ peak higher than the noise level (±0.04%).

The amount of attached HRP protein is difficult to determine from the intensity of the elemental $C_{1s}$, $O_{1s}$ and $N_{1s}$ peaks. The protein peaks overlap the same peaks of the PIII treated polyethylene. The penetration depth of XPS analysis is close to the thickness of the protein layer and the subtracted spectra cannot be used to consistently subtract the polyethylene peaks. The $S_{2p}$ region can be used for analysis as there is no sulphur in the polyethylene, however the concentration of sulphur atoms and corresponding intensity of $S_{2p}$ peak is low in the protein and gives quantitative measurements of limited accuracy. Following these results, we used XPS for qualitative analysis of protein on the surface as only additional support for the FTIR results.

### 3.4. Conformation of attached protein

Knowledge of the conformation of an attached protein is important to understand the activity of the protein molecule and its functionality in a biological system. Attached HRP protein on untreated hydrocarbon polymeric materials does not preserve its native conformation [94]. We have investigated the conformation of attached protein on PIII treated polyethylene using the Amide I line in FTIR ATR spectroscopy.

The spectra of HRP in the Amide I region were analysed for bulk HRP protein dried on Germanium crystal, and for attached HRP protein on PIII treated polyethylene surfaces. The bulk layer was selected as a control because the interactions of the protein molecules with any substrate are absent. The spectrum in the Amide I and II regions for the protein on PIII treated polyethylene is similar to the spectrum in the same regions for a bulk layer on a Ge crystal. The position of the maximum, the width and shape of lines remain the same when the protein is covalently attached.

The detailed analysis was done by fitting the Amide I region using individual peaks. These individual peaks characterise the presence of different structures in the protein molecule: α-helices, β-sheets and random chains. The individual conformation of a protein is characterised by unique number of concentrations of such structures. The change of conformation results in a change in the relative concentrations of these structures.

For a first approximation to determine the number and positions of individual peaks, the 2$^{nd}$ derivative of the measured spectrum and a deconvolution of it using Fourier transformation were evaluated and interpreted with the aid of the literature [95-97]. A simple unconstrained fitting procedure to extract the multiple components was unsatisfactory as the position and width of the components were not well determined. The fitting procedure was therefore carried out by constraining the width and positions of the individual peaks to a narrow range. The results of fitting are presented in Fig.15. The 1683 cm$^{-1}$ and 1676 cm$^{-1}$ lines are interpreted as vibrations in regions with β-turns, the 1661 and 1649 cm$^{-1}$ lines are attributed to vibrations in α-helices, the line at 1636 cm$^{-1}$ is attributed to random structures and the line at 1623 cm$^{-1}$ is interpreted as vibrations in β-sheets. Using equation (6), the relative concentrations of the protein chain structures were calculated from the intensity of the individual peaks.

Following the fitting results, the HRP protein in the bulk dry layer consists of 49% of α-helices, 23% β-turns, 17% of random structures and 11% of β-sheets. This structure is in agreement with the reference data of 50% of α-helices of HRP in crystal form determined from X-ray diffraction [98] and 30-40% of α-helices of HRP in PBS buffer solution [97]. The intensity of the spectra of the attached layer is significantly lower than in bulk protein layer, but the noise level is still low enough to get useful fitting results. The attached protein on an untreated surface has 45% of α-helices, 19% β-turns, 14% random structures and 17% β-sheets. The conformation of attached protein on the PIII treated surface was found to consist of 43% α-helices, 23% β-turns, 14% random structures and 12% β-sheets. These values are close to the values of HRP protein in the bulk layer. Therefore, we conclude that the conformation of HRP protein in a covalently attached monolayer on the PIII treated polyethylene surface remains similar to that in a bulk layer.

### 3.5. Catalytic activity of the attached protein

The attached protein is active in catalysing the decomposition reaction of peroxide. The activity is provided by interaction of a porphyrin ring with a calcium ion as its active centre with the peroxide molecule. When the conformation of protein has been changed, the activity of the protein is lost. Therefore, the analysis of the protein activity can be used as a simple test for the presence of HRP in its native conformation.

The activity of HRP can be tested with TMB and ABTS assays. The assay molecules are changed by reaction with oxygen ions, which come from the catalytically decomposed hydrogen peroxide. The changes in assay molecules are quantified using the colour of the solution as determined by an optical density in UV-vis spectroscopy. The rate of the decomposition reaction depends on the concentration of active protein molecules in solution. The optical density of the test solution can therefore be used as a measurement of protein activity. In the case of a high concentration of peroxide, the rate of the reaction follows the first order dependence on the concentration of the protein molecules:

$$dC = P \cdot dt \qquad (7)$$

where C is the concentration of coloured assay product molecules, P is the concentration of active protein attached on the surface.

The optical densities of TMB solutions, in which untreated and PIII treated polyethylene with attached protein are immersed, are presented in Fig.16a. The amount of free protein in solution was monitored and found to be small so that all changes in assay colour are caused by surface attached protein. The zero catalytic activity of the polyethylene before and after modification without attached protein was also confirmed.

For comparison of the activity of molecules in solution with those attached on surfaces, the activity of protein in solution was tested with the same assay. The concentration in solution was prepared by dilution of a known amount of protein corresponding to the 80 ng/cm$^2$ of protein on the surface at complete coverage [41]. During the dilution procedure, the walls of all glass containers were blocked by BSA to prevent loss of the protein due to attachment of the HRP on the glass walls.

The curve of the kinetics for the assay reaction of attached protein on untreated polyethylene is a factor of 2 lower than for attached protein on PIII treated polyethylene. Therefore, the amount of active protein on the PIII treated surface is a factor of 2 higher than for protein on the untreated surface. This difference corresponds to the difference in total amount of protein attached on the surface as confirmed by FTIR ATR spectra.

The kinetics curve of diluted protein in solution is higher than for attached protein on PIII treated surfaces. Because the HRP forms a monolayer on the polymer surface, the amount of active protein is close to the amount of total protein attached on the modified surface within experimental error. The difference between the activity of diluted protein in solution and protein attached on the surface can be explained by steric hindrance of the assay molecules in accessing the active centre of the protein. The presence of the polymer surface on one side and the dense packing of protein molecules on the surface both contribute to the steric hindrance. The influence of the density of packing on activity is shown in Fig.16b as a plot of protein activity as a function of the amount of protein attached as measured by FTIR ATR. The saturation in the curve shows this steric hindrance effect on the assay.

### 3.6. Stability of attached protein

Attached protein can be removed from untreated polyethylene surface with sufficient time of washing in buffer solution only. The kinetics of washing is presented in Fig.17 and is determined by the probability of breaking of the intermolecular interactions between the protein and the surface. The activity of the attached protein is proportional to the amount of protein.
Attached protein from an untreated polyethylene surface can be more quickly removed by washing in detergents. The detergents sodium dodecylsulfate (SDS), Triton X100 and Tween 20 were used for washing. These detergents are accepted as effective detergents to wash proteins away from different surfaces [99, 100]. Due to presence of polar and non-polar chemical groups in detergent molecule, these detergents provide a number of strong interactions with the protein and break the intermolecular interactions between the protein molecules and the polyethylene surface. Once the interactions are broken, the protein molecule is washed away. This process on an untreated polyethylene surface is observed by the disappearance of the protein peaks in FTIR spectra and the loss of activity of protein after washing in detergents.

Since part of the protein attached on a PIII treated polyethylene can be washed in buffer, this portion of the protein is not bonded covalently. However, some of the protein remains attached on a PIII treated polyethylene surface even after a long time of washing. The same amount of protein remains on PIII treated protein after washing in SDS, Triton X100 and Tween 20 detergents. These protein molecules cannot be washed away because of strong interactions between protein molecule and PIII modified surface. Such strong interactions cannot be provided by physical intermolecular interactions, but can be explained by the formation of chemical bonds between protein molecules and the surface. These chemical bonds cannot be broken with these detergents even after long washing times.

It has been shown previously that the attachment of proteins depend on the pH its environment (buffer) [54, 101]. The pH effect on attachment of HRP was tested experimentally. The protein from buffer solution with various values of pH was attached on untreated and PIII treated polyethylene surface. The amount of attached protein was measured by FTIR ATR spectroscopy and was found not to depend on the pH of the buffer for both untreated and treated polyethylene as shown in Fig.18a. The amount of covalently attached protein after washing in SDS detergent was also measured by FTIR ATR spectroscopy and found not to depend on the pH of the soaking solution over the range pH6 to pH8.

In general, the addition of ionic compounds into a buffer changes the ionic force of the buffer and can change the absorption of protein onto a surface immersed in the buffer [54, 101, 102]. The influence of adding NaCl on HRP protein attachment was tested experimentally. The protein from solution with different concentrations of NaCl was attached on untreated and PIII treated polyethylene surfaces. The amount of attached and covalently bonded protein does not depend on the salt concentration in the range 0 to 1 mMol, as shown by FTIR ATR spectroscopy (Fig.18b). The same result was observed for various concentrations of $Mg(OH)_2$ additive in HRP solution.

The concentration dependence of the amount of HRP attached on treated and untreated polyethylene in buffer after overnight incubation is shown in (Fig.18c). In the low concentration region (up to 10 mg/ml), the amount of attached protein on untreated polyethylene increases with protein concentration. This can be explained by an increasing probability of protein contact with the polyethylene surface. In the region of high concentrations (from 10 mg/ml of protein), the amount of attached protein on untreated polyethylene does not depend on the protein concentration in solution. The amount of protein attached on PIII treated surface is higher than attached on untreated polyethylene for all applied concentrations and does not depend on the concentration of the protein in solution.

After washing in detergent, the amount of protein attached on untreated polyethylene was significantly decreased and remains constant for all applied concentrations of protein in solution. The amount of protein attached on PIII treated polyethylene does not change (in error bar diapason) after washing in detergent for protein solutions with low concentration (up to 10 mg/l). For solutions with higher concentration (from 10 mg/l), some protein attached on the PIII modified surface was washed away. However, the amount of covalently bounded protein on PIII treated surface remains independent of the concentration of protein solution. Therefore, the attachment of protein on untreated and PIII treated surfaces behaves differently.

The catalytic activity of the HRP is connected with its specific conformation. The influence of conformation change (unfolding) on the attachment process was analysed by unfolding the HRP prior to attachment. The unfolding of the HRP was done with boiling in buffer solution for 30 minutes after which all activity was destroyed. After that the protein was attached on PIII treated and untreated polyethylene surfaces. The amount of the attached and covalently attached inactive protein was found to be similar as for active protein. Therefore, the conformation of HRP protein does not play role in the attachment process on either untreated or PIII treated polyethylene surface.

The amount of attached protein of PIII treatment increases with time and shows saturation (Fig.18d). The amount of covalently attached protein increases with the time of PIII treatment and also shows saturation. The concentration of free radicals increases in the same range of PIII treatment times. A perfect correlation between the amount of attached protein and the measured free radical concentration is not observed because the measured free radical concentration includes radicals in the bulk of the polyethylene. Only surface free radicals are effective in the attachment process. This difference in roles of the two types of radicals has the effect of a saturation in the protein amount earlier than the saturation of the free radical concentration. The saturation of the amount of protein covalently attached is achieved after about $10^{15}$ ions/cm$^2$ fluence of ion beam.

The ability to attach a protein with chemical bonds to PIII treated polyethylene remains after long times. To investigate the aging effect of PIII treatment, the polyethylene samples were treated and stored under laboratory conditions (in a closed dark container, stabilized at a temperature of $23^0$ C and at 70 % relative humidity) for two years. The amount of attached protein on freshly prepared PIII surface is slightly higher than on the aged PIII treated surface (Table 1). The amount of covalently bonded protein on a 2 year old PIII treated polyethylene surface and on a freshly treated surface is the same within error bars. Together with protein attachment, the ATR FTIR spectra and the surface energy of PIII treated samples do not show significant changes during two years of storage after a stabilization in first month after PIII treatment. Therefore, the surface remains active to covalent attachment of protein after at least two years of storage.

### 3.7. Functionalization of the polyethylene surface and protein attachment/ Influence of oxygen containing groups on covalent attachment of protein

In this section we analyse previous explanations of the attachment of protein on surfaces that have been treated by ions, electrons or ionising radiation. If the attachment of protein on a PIII treated surface is provided by chemical bonds, which are formed between protein molecules and surface active groups during incubation of the treated polymer in a protein solution, the surface activity can be simulated by artificially introducing active groups into the surface layer during synthesis of the polymer film. The conditions of incubation are not specific as described above, therefore the reaction (or reactions) must be realised easy. It means that the groups of modified layer must be active enough to take part in chemical reaction with protein under normal conditions of soaking. The untreated polyethylene does not provide the chemical bonds, the active groups in the surface of polyethylene appear after modification, therefore the analysis of new groups appeared after PIII can be used for creation of simulated active surface layer to protein attachment.

To investigate the groups responsible for protein attachment, the modified polyethylene surface was simulated by introducing selected chemically active groups. It is well documented that oxygen containing groups such as carbonyl, carboxyl, aldehyde and hydroxyl can provide chemical bonds with protein molecules [54]. As described above, the PIII modified layer of polyethylene contains a number of oxygen containing groups that appear as the result of free radical reactions with atmospheric oxygen after exposure of the treated samples in air. The presence of such active groups is considered in the literature as a reason for the strong adhesion of modified polyethylene surfaces to epoxy and polyurethane adhesives. The reactions responsible are between epoxy and isocyanate groups of the adhesive with carboxyl and hydroxyl groups of the modified polyethylene [71] in the interface region. Similar reactions can be expected to be responsible for the covalent protein attachment.

The presence of hydroxyl, carboxyl and aldehyde groups on the polyethylene surface after PIII would give rise to a potential for a reaction between the amine group of Lysine and these oxygen containing groups:

$$R\text{-}OH + H_2N\text{-}Protein \rightarrow R\text{-}NH\text{-}Protein + H_2O$$

$$R\text{-}COOH + H_2N\text{-}Protein \rightarrow R\text{-}CO\text{-}NH\text{-}Protein + H_2O$$

$$R\text{-}COH + H_2N\text{-}Protein \rightarrow R\text{-}CH\text{=}N\text{-}Protein + H_2O$$

These reactions are used to bind HRP to specific compounds in a number of biotechnological tools [3, 5, 54, 95, 102].

The reaction of carboxyl group is more probable due to higher activity of carboxyl group with amine than the activity of hydroxyl or aldehyde groups with amine. The carboxyl groups can be introduced in a polymer surface by means of co-polymerisation reactions without any need for radiation damage to the polyethylene. To prove the concept of using carboxyl-amine reactions for protein immobilisation we intentionally introduced active carboxyl groups into polyethylene.

The co-polymer of ethylene and methacrilyc acid with 17 wt% concentration of methacrylic acid monomer was used as a model surface. The MMA concentration corresponds to the theoretical density of 1 carboxyl group per 14 methylene groups in the co-polymer. The samples were moulded between hot plates covered by Teflon film from granules of co-polymer. After moulding, the surface of the co-polymer was treated with ethanol, acetone and water. The concentration of residual Teflon macromolecules on the co-polymer surface after moulding was monitored by ATR-FTIR spectra and found to be undetectable. The ratio of the $\nu(C=O)=1690$ cm$^{-1}$ from the carboxyl group and the $\nu(CH_2)=2917$ cm$^{-1}$ lines of the polyethylene absorbance was found to depend on solvent treatment (ethanol, acetone and water) after moulding. The hydrophilic solvent causes slightly higher concentration of carboxyl groups in the surface layer observed as a higher intensity of $\nu(C=O)$ line than freshly moulded or after hydrophobic solvent treatment (Table 2). However, even after treatment with the hydrophobic solvent toluene, the surface layer of the moulded film still contains carboxyl groups, although somewhat reduced in concentration. The activity of carboxyl groups was confirmed by a test reaction with sodium hydroxide which causes the appearance of a sodium carboxylate line at 1574 cm$^{-1}$. Therefore, the carboxyl groups on the surface of the moulded film are active. The surface of polyethylene with introduced carboxyl groups becomes slightly hydrophilic, the water contact angle decreasing from $90^0$ for the as received polyethylene to $88^0$ for the co-polymer.

The activated polyethylene samples were soaked in protein solution as described above and tested by FTIR ATR spectra and protein activity. The amount of attached HRP molecules on the surface with carboxyl groups is lower than on untreated polyethylene surface. After washing with detergents, the amount of HRP molecules decreased below the level for untreated polyethylene. Therefore, the presence of active carboxylic groups in a co-polymer does not provide a covalent attachment of HRP.

The minimisation of the influence of oxygen-containing groups in PIII treated samples could be realised in an experiment in which a protein is attached on the surface that has never had contact with air after the PIII treatment. In a realisation of this experiment, the protein solution was placed in a vacuum chamber termed the "chemical vacuum chamber", separate but connected to, the plasma treatment chamber and this chamber was pumped down to about 20 Torr pressure when water starts to boil. The chemical chamber was then vented with nitrogen up to atmospheric pressure. The procedure of pumping and venting was repeated 5 times. Then the container with protein solution was closed and the chemical vacuum chamber was pumped down to $10^{-3}$ mTorr.

The PIII treated sample of polyethylene was moved from the plasma chamber into the chemical chamber under a vacuum of $10^{-3}$ mTorr by means of an arm manipulator. Then the chemical chamber was isolated from the plasma chamber, vented with pure nitrogen up to either a partial vacuum (20 mTorr) or 760 Torr and the container with protein solution was opened into the chemical chamber. The sample was immersed in the protein solution and soaked there overnight. After that, the chamber was vented and opened to air and a washing procedure to remove loosely bound protein was applied. Due to the exclusion of oxygen from the freshly treated PIII treated sample until protein was applied, the oxygen-containing group concentration is expected to be significantly less than in the normal process in which samples are exposed to air after PIII treatment (see section 3.1) and the concentration of free radicals is expected to be higher.

The amount of attached protein in partial vacuum (20 mTorr of nitrogen) and in nitrogen atmosphere (760 Torr) was found to be as high as for a PIII treated sample exposed to air as usual. The attached protein remains on these surfaces after all detergent washing, showing that it is covalently bound.

These experiments confirm the copolymer experiments described above and confirm that the presence of carboxyl groups on the surface is not responsible for protein attachment and cannot cause covalent bonding of protein by means of carboxyl groups on the surface reacting with an amine group of the protein.

The activity of other groups such as aldehyde, hydroxide, lactone, carbonyl and others is less than for carboxyl and the reactions of these groups with amine groups of aminoacid residuals has a smaller probability. For example, the attachment of HRP molecules on PMMA with a high concentration of ester groups, on polyamide with high concentration of amide groups, on PEEK with high concentration of ester and carbonyl groups, is lower or the same as for untreated polyethylene [42, 103]. The effect of carbon-carbon unsaturated groups on protein attachment can be studied with polystyrene which contains aromatic rings. As shown above, the attachment of protein on untreated polystyrene is low [41]. A full analysis of the active groups present in a selection of representative polymers is shown in Table 3.

Therefore, the presence of active oxygen containing and unsaturated groups in the surface layer cannot cause high attachment of the protein as well as covalent bonding of protein and the appearance of such groups in the surface layer of PIII modified polyethylene is not the reason of high attachment of protein and covalent bonding of protein molecules to the surface (Table 3).

The surface of PIII modified polyethylene contains free radicals with a high concentration. The free radicals disappear over time due to their high activity in accepting any mobile atom or molecule capable of reacting with it from the environment. Free radicals have the ability to migrate along macromolecules and between macromolecular chains from the surface layer into the bulk. At the same time, free radicals are stabilized by the presence of condensed aromatic rings due to the delocalisation of electrons on pi orbitals in aromatic structures. Also Free radicals are also stabilized by the presence of nitroxil groups and thiazyl groups. The stabilization of free radicals is increased by steric hindrance effects. This applies when the free radical belongs to a group, which is surrounded by neighbouring groups in such a way that any other molecule cannot come close to the free radical. All these mechanisms are discussed in the literature [83-85]. In our experiments on several polymers, the ESR signal of free radicals remains for a long time (months) after PIII treatment.

Active free radicals are present on the surface of polymers after PIII treatment. This was proved in previous experiments on ion beam modified PTFE [104], when the free radicals from modified PTFE caused crosslinking reaction in thick layer of a resin placed on the PTFE surface. The free radicals reactions on modified polymer surface have been proposed as a mechanism for protein attachment [105].

Free radicals can take part in the chemical reactions with protein molecules and provide the formation of chemical bonds between the modified surface layer of a polymer and an attached protein. However, a synthesis of initial polymer macromolecules with active free radicals in the surface layer appears not to be possible.

FTIR analysis of the quantity of HRP showed that the amount of protein attached to a sample exposed only to vacuum is the same as for air-exposed sample (Table 4). However, the activity of protein is less in a sample exposed in vacuum, than for the air-exposed sample. The deconvolution of the Amide I line shows that the conformation is different for protein attached to a surface in air and in vacuum (Fig.19). The intensity of 1638 cm$^{-1}$ line, attributed to disordered chains, is higher for the vacuum attached protein. In addition, the spectrum to spectrum deviation for different samples is stronger for samples that had protein attached in vacuum. Such a strong deviation implies that the protein conformation is random. Such random conformation variations suggest that the conformation is subject to change by a random factor such as a high concentration of free radicals randomly distributed on the surface. A high concentration of free radicals will lead to more than one covalent bond forming between the molecule and the surface. This multiple bond formation will have the effect of causing both the random conformation changes and the loss of protein activity through excessive unfolding.

In summary, there is a significant body of evidence that free radicals are responsible for the formation of the observed covalent bonds between protein and a PIII modified surface.

**3.8. Blocking of covalent attachment to PIII treated surfaces**

If free radicals are responsible for covalent attachment, it should be possible to block the activity of free radicals selectively by using specific blocking agents that are applied after the PIII treatment and before the protein attachment.

It was found that covalent attachment of protein on PIII treated polyethylene can be blocked by the application of specific low molecular weight components. In experiment, the PIII modified polyethylene surface was treated by Styrene, Hexene, polytetrahydrofurane (PTHF), Ethylenediamine, Benzyl mercaptan and 2,2,6,6-Tetramethylpiperidine 1-oxyl (TEMPO) in different solvents (toluene, acetone and ethanol). After treatment, the polyethylene samples were washed with pure solvents to remove non-bonded blocking molecules. The presence of the applied blocking agents on the polyethylene surface was monitored by FTIR ATR spectra using specific group lines of the attached molecules. The control experiment with initial polyethylene showed that the blocking molecules do not stick to the unmodified surface and the washing procedure is sufficient to remove all unbounded blocking molecules. All blocking agents are observed as attached to the PIII treated surface. After blocking, the protein was attached on the treated polyethylene surface as described above. The amount of attached protein before and after SDS detergent washing by FTIR ATR spectra are shown in Table 5.

The amount of attached protein depends on the details of the intermolecular interactions between protein molecules and the blocked surface, which in turn depends on the blocking agent. The adsorption of hydrocarbon compounds (hexene and styrene) on a PIII modified surface does not change the amount of attached protein very much. The presence of polar groups in the blocking molecule (for example, ethylenediamine, polytetrahydrofurane and TEMPO) decreases the amount of the attached protein. This trend is in accordance with known data on a protein attachment on hydrophobic surfaces [49].

The amount of covalently attached protein, as tested by SDS washing, is the same on adsorbed hexene and styrene layers as it is on the unblocked PIII treated surface. The amount of covalently attached protein decreases on surfaces blocked with ethylenediamine and PTHF. The lowest amount of covalently bonded protein is observed on the surfaces blocked with adsorbed TEMPO and Benzyl mercaptan.

None of the blocking molecules tested are expected to have a specific chemical activity to the protein once attached on the surface and the chemical attachment of the protein molecules to the blocked surface is therefore not expected. Blocking agents that contain free radicals may show a different behaviour on a surface that contains free radicals. Hexene and styrene molecules contain carbon double bonds which open in the presence of free radicals to give a possibility of polymerisation to form a polyhexene or polystyrene layer:

$$R\text{-}CH=CH_2 + \bullet R' \rightarrow R\text{-}CHR'\text{-}CH_2\bullet$$

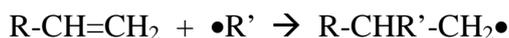

The final radical on the right hand side can be used for the next reaction with a new molecule of styrene or hexene or the radical could be kept and used for the protein attachment. Also, one layer of polystyrene or polyhexene macromolecules synthesised on the surface is able to transfer a free radical along the macromolecule to the new surface above the attached layer:

$$R\text{-}CH\bullet\text{-}CH_2\text{-}R' \rightarrow R\text{-}CH_2\text{-}CH\bullet\text{-}R'$$

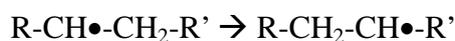

Following this scheme, the hexene and styrene cannot block the covalent attachment of the protein, in agreement with experimental observation.

Ethylenediamine and PTHF contain oxygen or nitrogen in backbone chain of the molecule. In the presence of free radicals, these molecules can be attached by the hydrocarbon part of the molecule to the surface, but the oxygen and nitrogen atoms in these molecules are likely to be separated from the parent molecule in the presence of a free radical and the free radical cannot be kept in the blocking molecule.

$$R\text{-}CH\bullet\text{-}CH_2\text{-}O\text{-}R' \rightarrow R\text{-}CH=CH_2 + \bullet O\text{-}R'$$

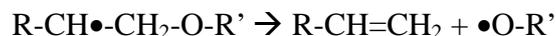

Therefore, these molecules cannot transfer a free radical through the layer. Therefore, after attachment of these molecules the protein attachment becomes impossible. These two molecules (Ethylenediamine and PTHF) can partially block the surface against the protein attachment, which is observed in experiment.

The highest blocking ability is observed for TEMPO and benzyl mercaptan molecules. TEMPO contains a free radical stabilized by aromatic ring conjugated with nitroxyl group, but the activity of the free radical in TEMPO is high enough to cause free radical reactions in chemical processing. Therefore, the attachment of TEMPO to PIII modified surface is expected to be a reaction between two free radicals – one from TEMPO and one from the surface:

$$R\bullet + \bullet O\text{-}R' \rightarrow R\text{-}O\text{-}R'$$

After attachment of TEMPO the PIII modified surface does not have free radicals and the protein attachment is blocked completely. The same reason is with benzyl mercaptan molecules, where S-H group is very active to free radicals:

$$2R\bullet + HS\text{-}R' \rightarrow R\text{-}H + R\text{-}S\text{-}R'$$

As the result, the free radicals on the surface react with benzyl mercaptan and TEMPO that renders the modified polyethylene surface to be inactive for protein attachment.
Therefore, the reactions of specific blockers support the free radical mechanism for the covalent binding of protein to PIII modified surfaces.

### 3.9. Attachment of polyaminoacids

This study was undertaken to determine which chemical groups in a protein take part in the covalent bonding to a PIII modified surface via a free radical. Free radicals are able to react with a wide range of molecules. From this point of view, it is likely that all aminoacid residues in a protein molecule can take part in covalent bonding with a PIII treated surface. One chemical bond is enough to give a covalent attachment of protein molecule. To confirm that the covalent binding is possible via many different types of aminoacid residues, we attached a number of different polyaminoacids on PIII treated polyethylene. In this way, we simulated by attachment of selected aminoacid residues when they are present in a protein molecule.

The activity of polyethylene surface after ion beam implantation to free alanine and leucine was observed earlier [106]. But due to their low molecular mass, it is possible that individual aminoacids could diffuse into the bulk. To confirm that the reaction is on the surface, we used polyaminoacids of high molecular mass to exclude diffusion into the bulk. In 20 of the most common aminoacids, there are 7 different chemically active groups present in the side chains (methyl, aromatic ring, hydroxyl, amine, mercapto, carboxyl and pyridine). Therefore, the activity of 20 aminoacids can be simulated by attachment of 7 different polyaminoacids, each having one of the 7 different side chain groups.

The attachment of polyaminoacids was carried out on untreated and PIII treated polyethylene and tested with FTIR ATR spectroscopy. The molecular mass of polyaminoacids was selected as to be as close as possible to the molecular mass of HRP. All polyaminoacids were attached from aqueous buffer solution, except Poly-l-tryptophan which has limited solubility in aqueous solution and was attached from acetone solution. The SDS washing procedure to test for covalent attachment was the same as for protein attachment except for Poly-l-tryptophan, which was washed with pure solvent only. The results of attachment experiments are presented in Table 6.

As can be seen from the Table, not all polyaminoacids are well attached. This means that the end-groups (base and acid) and the amide backbone groups presented in all aminoacids can be excluded from consideration as active binding groups, and the binding must take place via a side group.

An amount of attached Poly-l-histidine and Poly-l-lysine is high both before and after SDS washing. This is similar to the behaviour observed for HRP on the PIII treated polyethylene. Poly-l-glycine is a hydrophobic polyaminoacid and the amount of Poly-l-glycine present both before and after washing is lower than the amount of attached Poly-l-histidine and Poly-l-lysine. This expected since the PIII treated polyethylene is mildly hydrophilic. However, some Poly-l-glycine molecules remain after detergent washing on the PIII treated surface, showing that covalent bonding takes place.

Poly-l-tryptophan, Poly-L-methionine, Poly-L-threonine, Poly-L-isoleucine and Poly-DL-alanine were attached from acetone solution. In his case, the physical interactions between the surface and these molecules are quite different from those that would apply in aqueous solution due to presence of acetone molecules. However the attachment is provided by chemical bonds on PIII treated surface should be the same, independent of the environment. The chemical bonding will be covalent since no physical interactions will be strong enough to withstand washing in pure acetone.

Poly-l-glutamic acid has a low attachment ability on the PIII treated surface. The PIII treated polyethylene has acid groups on the surface as discussed in section 3.1. In buffer solution the acid groups are converted to ionic form and prevent close contact of the surface with acidic molecules from solution. Therefore, it is likely that Poly-l-glutamic acid molecules cannot come close enough to the surface to allow a chemical reaction with free radicals.

Poly-l-tyrosine was attached from a basic solution due to its low solubility in a pH neutral solution. Because of the presence of carboxyl acid groups on the surface and sodium ions in solution, the Poly-l-tyrosine molecules have the same problem as Poly-l-glutamic acid to achieve a sufficiently close interaction with the surface to achieve any kind of chemical interaction. This could be the reason why Poly-l-tyrosine shows low attachment ability on the PIII treated surface. Note however, that Poly-l-tyrosine shows high attachment ability on the untreated polyethylene surface where ionic interactions near the surface do not play a major role. Besides that, Poly-L-threonine is strongly attached from acetone solution.

A mercapto group reaction with PIII treated polyethylene surface was observed at poly-L-Methionine attachment. Methionine and Cystein -SH groups are expected to be involved in chemical reaction with free radicals of PIII modified surface.

In summary, all polyaminoacids are potentially capable of attachment to the PIII treated surface. The low attachment that was observed in specific cases we explain by specific ionic interactions in the layer of buffer near the surface. Such a universal behaviour of aminoacids residues with the PIII modified surface is provided by the presence of active free radicals. However, in a protein molecule, the ability of an individual aminoacid residue to take part in chemical bonding with the PIII treated surface depends also on the conformation of the protein and on the orientation of the protein molecule relative to the surface.

The universal character of protein attachment to free radicals could be extended to various methods of radiation damage of polymers, if it can be demonstrated that these methods provide a sufficient concentration of active free radicals on or near the surface. In that case the mechanism of protein attachment described in this work could be applied to damage by glow discharge plasma, UV light irradiation, electron beam modification, X-ray modification, γ-irradiation and mechano-destruction (polishing and scratching) of carbon-containing polymers as well carbon deposition methods that create carbon structures (such as graphite clusters, carbon nanotubes and graphene) with stabilised free radicals. The stages of PIII treatment of polyethylene and protein attachment are presented in Fig.20.

## 4. Conclusion

Plasma immersion ion implantation is a type of ion beam implantation and produces structural changes in a thin surface layer of polyethylene. These changes include:
- a carbonisation process in which unsaturated groups containing double bonds, conjugated double bonds, aromatic rings, condensed aromatic rings in increasing numbers that eventually form graphitic islands;
- appearance of active free radicals and free radicals that show long-term stability;
- appearance of oxygen-containing groups such as carboxyl, carbonyl, aldehyde and hydroxyl groups after exposure of the treated sample to air;
- appearance of nitrogen-containing groups during implantation and after exposure of the treated sample to air;
- an increase in surface energy that coincides with the appearance of free radicals and decreases with storage time as the free radicals decay and oxygen-containing groups appear;
- a wave-like morphology of the surface.

The amount of protein attached on a PIII treated surface is up to 2 times higher than on an untreated surface. The attachment on the PIII treated surface is through the formation of covalent bonds between the protein molecule and the modified surface, while the attachment on the untreated surface is through physical intermolecular interactions.

The covalent bonds between the protein molecule and the modified surface are formed by a reaction with active free radicals, which are preserved in conjugated structures due to delocalisation of unpaired free electrons. The blocking of the modified surface with selected molecules eliminates free radicals or prevents their migration, in turn preventing the covalent attachment of protein.

The protein molecule has a number of active aminoacid residues available for bonding to free radicals at the surface. This means that the PIII treated surface can provide a covalent attachment for all kinds of proteins provided that the protein molecule is able to come close enough to establish a chemical bond. The universal character of protein attachment to free radicals could be extended to various methods of radiation damage of polymers, if it can be demonstrated that these methods provide a sufficient concentration of active free radicals on or near the surface.

The native conformation of attached protein is retained on a PIII treated surface due to hydrophilic interactions at the interface. However, a high concentration of free radicals on the surface can give rise to multiple covalent bonds with the protein molecule that destroy the native conformation with a consequent loss of catalytic activity. Therefore, the retention of protein conformation (and activity) requires an optimisation of the free radical concentration on the surface.

**Acknowledgements**

Authors would like to acknowledge the contribution of Helen Smith and Robert Thompson, who took part in experiments with polyaminoacids and proteins.

Table 1. Amount of attached HRP protein (determined from FTIR ATR using equation (6)) on UHMWPE treated by PIII with 20 keV energy Nitrogen ions at $10^{16}$ ions/cm$^2$ fluence.

| Substrate | As attached | Washed in SDS |
|---|---|---|
| Untreated | 0.0093±0.0014 | 0.0005±0.0007 |
| PIII treated, new | 0.0171±0.0001 | 0.0095±0.0001 |
| PIII treated, after 2 years | 0.0136±0.0004 | 0.0085±0.0021 |

Table 2. Amount of HRP attached protein (see equation (6)) on untreated and PIII treated UHMWPE and on co-polymer of polyethylene and methacrilyc acid (co-PE-MMA) before and after washing in detergents. The HRP protein attachment was done on PIII treated sample in air, in Nitrogen at low pressure (30 Torr) and high pressure (760 torr). Carbonyl group relative concentration is measured by intensity ratio of 1691 cm$^{-1}$ line and 1462 cm$^{-1}$ line in FTIR ATR spectra. The average error bar for the protein amount is 0.0010.

| Sample description | Amount of attached after incubation | Amount after SDS | Amount after Triton | Amount after Tween 20 | C=O concentration before attachment |
|---|---|---|---|---|---|
| Untreated | 0.0087 | 0.0041 | 0.0018 | 0.0023 | ~0 |
| PIII, in air | 0.0198 | 0.0143 | 0.0142 | 0.0167 | 0.14 |
| PIII, in N$_2$, 30 torr | 0.0182 | 0.0161 | 0.0106 | 0.0100 | ~0 |
| PIII, in N$_2$, 760 torr | 0.0255 | 0.0147 | 0.0186 | 0.0180 | ~0 |
| co-PE-MMA, pressed | 0.0086 | 0 | 0.0033 | 0.0024 | 0.46 |
| co-PE-MMA, pressed and ethanol washed | 0.0057 | 0 | 0.0014 | 0.0029 | 0.53 |

Table 3. Chemical groups and polymer representatives available for covalent attachment of protein.

| Active chemical group | Polymer representative | Is a protein covalently attached? |
|---|---|---|
| Methylene, -CH$_2$- | Polyethylene | No |
| Double carbon bond, -C=C- | Polystyrene | No |
| Hydroxyl, -OH | Polyvinylalcohol | No |
| Ether, C-O | Polyethyleneterephtalate | No |
| Carbonyl, C=O | Polyethyleneterephtalate | No |
| Aldehyde, CH=O | Polyacrolein | No |
| Carboxyl, -COOH | Polymethacrylic acid | No |
| Free radical, R• | No known example | ? |
| Combination of all these groups | PIII treatment of polyethylene | Yes |

Table 4. Amount (see equation (6)) and catalytic activity of HRP protein on PIII modified LDPE attached in air and in vacuum.

|  | attachment on air | attachment in vacuum |
|---|---|---|
| Protein amount as attached | 0.020±0.001 | 0.023±0.001 |
| Activity of protein | 0.54±0.06 | 0.45±0.02 |

Table 5. Amount of attached HRP protein (see equation (6)) on PIII treated UHMWPE with various blocking agents of the surface. Detergent SDS was applied for analysis of covalently attached protein. The average error bar for the protein amount is 0.0010.

| Blocking agent | Formula | As attached | After SDS washing |
|---|---|---|---|
| No blocking |  | 0.0164 | 0.0113 |
| Hexene | $CH_3(CH_2)_3CH=CH_2$ | 0.0162 | 0.0090 |
| Styrene | 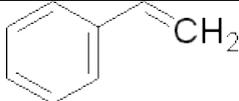 | 0.0156 | 0.0098 |
| Ethylenediamine | 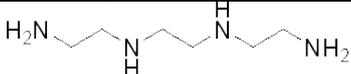 | 0.0122 | 0.0049 |
| Polytetrahydrofurane (PTHF) | 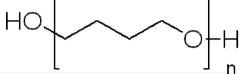 | 0.0102 | 0.0060 |
| TEMPO | 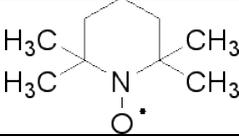 | 0.0119 | 0.0015 |
| Benzyl mercaptan | 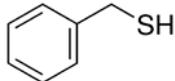 | 0.0078 | 0.0018 |

Table 6. Amount of Polyaminoacids (see equation (6)) on PIII treated polyethylene immediately after incubation (as attached) and after SDS washing (except for Poly-l-tryptophan, Poly-L-arginine, Poly-DL-alanine, Poly-L-isoleucine as described in the text). The average error bar for the polyaminoacid amount is 0.0005.

| Aminoacid | Polyaminoacid | Side-chain groups | As attached | After detergent washing |
|---|---|---|---|---|
| Glycine<br>Alanine<br>Isoleucine<br><br>Similar to:<br>Leucine<br>Valine<br>Proline | Poly-L-glycine (in buffer)<br>Poly-DL-alanine (in acetone)<br>Poly-L-isoleucine (in acetone) | $-CH_2-$<br>$-CH_3$ | 0.0092<br>0.0074<br>0.016 | 0.0025<br>0.0074<br>0.016 |
| Lysine<br>Arginine | Poly-l-lysine (in buffer)<br>Poly-L-arginine (in DMFA) | $-CH_2-$<br>$-NH_2$ | 0.0077<br>0.014 | 0.0073<br>0.014 |
| Histidine | Poly-l-histidine (in buffer) | $-NH-$<br>$=N-$<br>$-CH=$ | 0.016 | 0.013 |
| Tryptophan | Poly-l-tryptophan (in acetone) | $-NH-$<br>$-CH=$ | 0.0171 | 0.0087 |
| Tyrosine<br>Threonine | Poly-l-tyrosine (in buffer)<br>Poly-L-threonine (in acetone) | $-OH$<br>$-CH=$ | 0.0022<br>0.011 | 0.0002<br>0.011 |
| Glutamic acid<br><br>Similar to:<br>Aspartic acid | Poly-l-glutamic acid (in buffer) | $-CH_2-$<br>$-COOH$ | 0.0002 | 0.0002 |
| Methionine<br><br>Similar to:<br>Cystein | Poly-L-methionine (in acetone) | $-CH_2-$<br>$-SH$ | 0.025 | 0.025 |

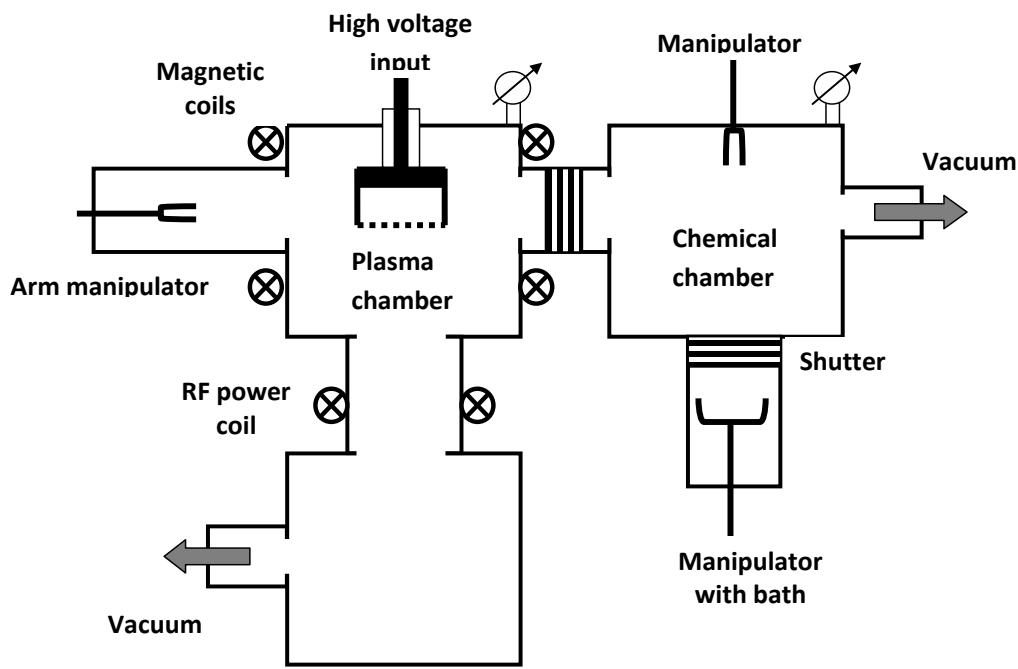

Fig. 1. Plasma immersion ion implantation chamber with attached chemical chamber for post-treatment of polymer sample after PIII treatment. The sample is treated in "plasma chamber" by plasma and ion beam, which is extracted from plasma under high voltage. The high energy ions are bombarding the sample through the mesh. After treatment the sample is moving with the arm manipulator into "chemical chamber", where the top manipulator grabs the sample and put to the bath with solution. The manipulator with bath can be replaced with ATR crystal when FTIR ATR spectra of PIII treated sample can be recorded under vacuum. All manipulations can be done under vacuum or inert gas.

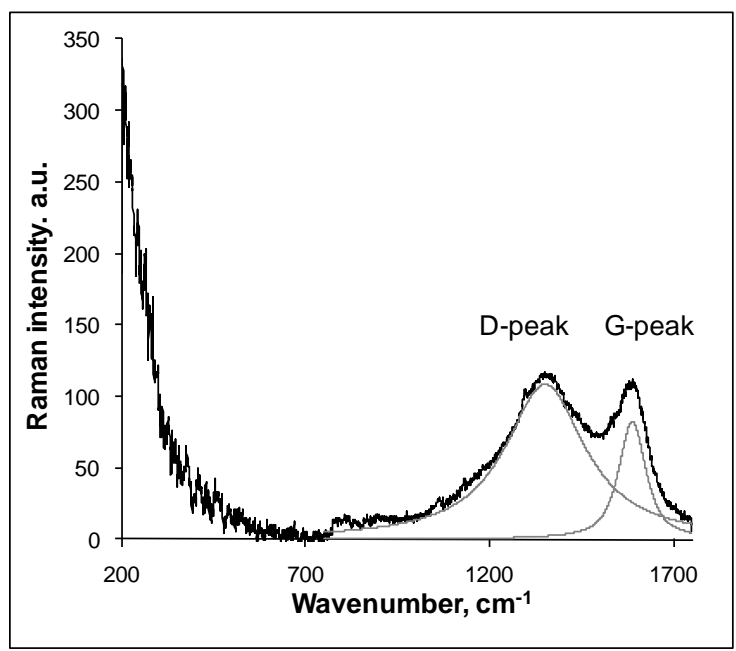

Fig.2. Micro-Raman spectrum of polyethylene treated by PIII with nitrogen ions of 20 keV energy and $10^{16}$ ions/cm$^2$ fluence. The spectrum is fitted with mixed Lorentz+Gauss functions at 1350 and 1587 cm$^{-1}$ positions corresponding to the "D" and "G" peaks of carbon structures respectively.

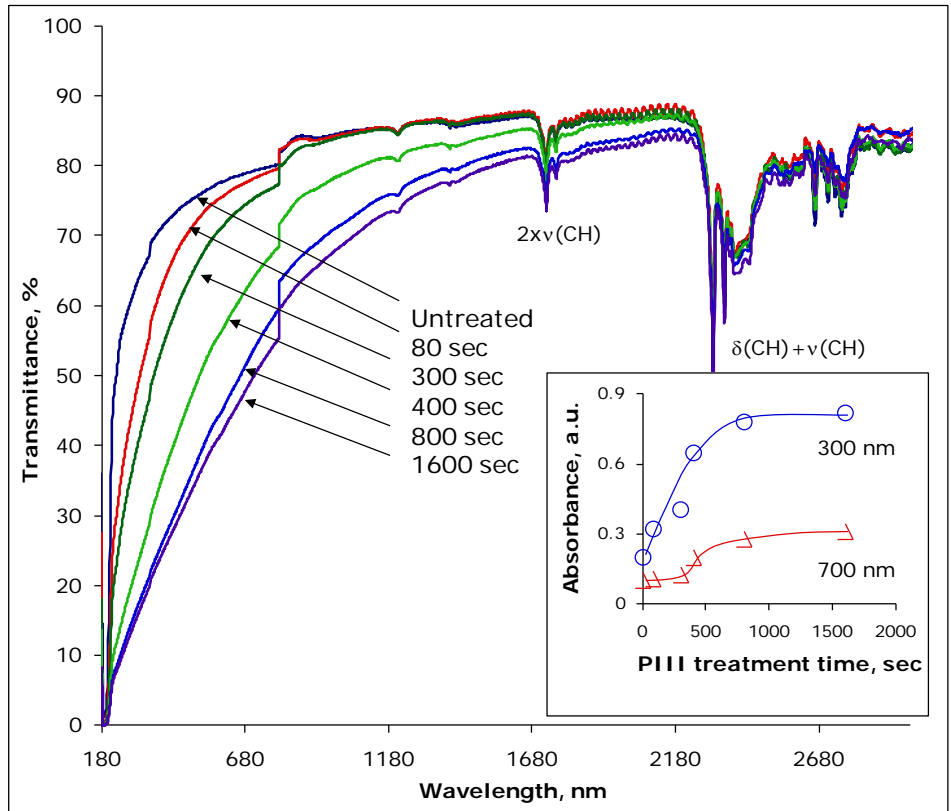

Fig.3. UV-Vis transmission spectra of polyethylene film after PIII treatment to a range of fluences of nitrogen ions with 20 keV energy. The time of PIII treatment is shown on the graph. The absorbance at short wavelength (300 nm) grows quicker with PIII treatment time, than the absorbance at long wavelength (700 nm).

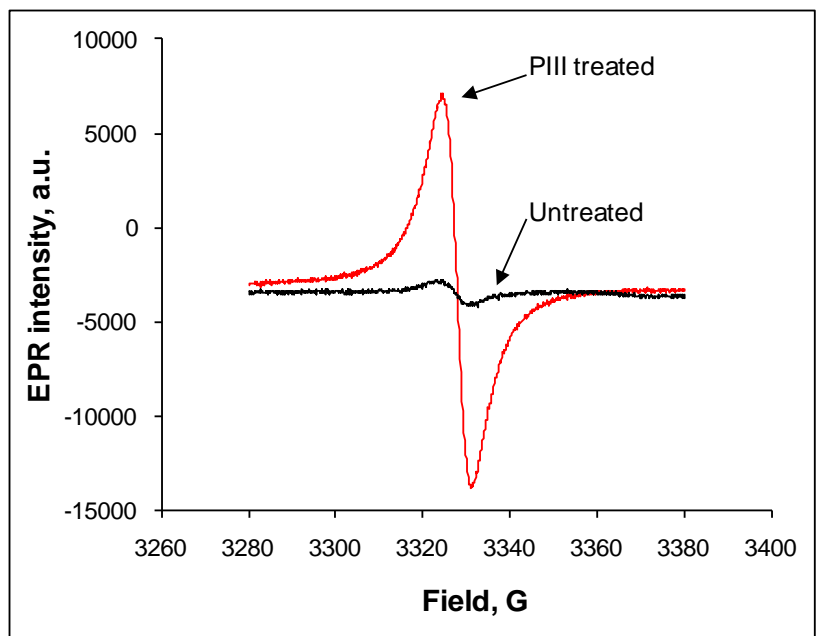

Fig.4. ESR spectra of LDPE film before and after PIII treatment. PIII treatment was done with 20 keV energy nitrogen ions to an ion fluence of $2 \times 10^{16}$ ions/cm$^2$. The spectrum of the treated polyethylene surface was recorded after 30 minutes of exposure to air at room temperature.

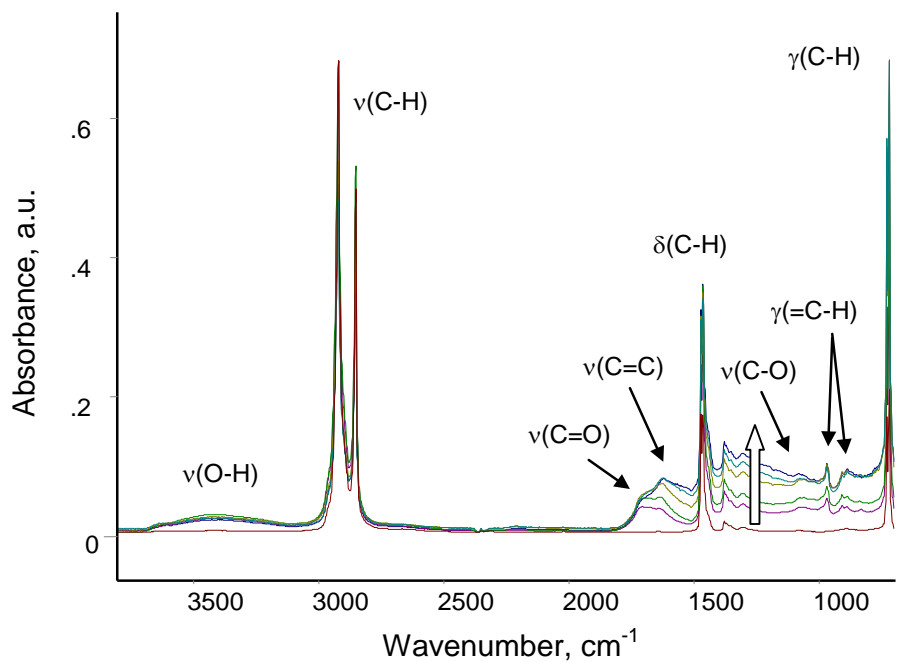

Fig.5. FTIR ATR spectra of LDPE after PIII treatment. Thick array shows fluence increase of PIII treatment: from untreated to $10^{15}$, $2 \times 10^{15}$, $5 \times 10^{15}$, $10^{16}$, $2 \times 10^{16}$ ions/cm$^2$. The spectra were recorded after stabilization during 2 weeks storage time after PIII treatment. The intensity of unsaturated and oxygen-containing group lines increases with fluence of PIII.

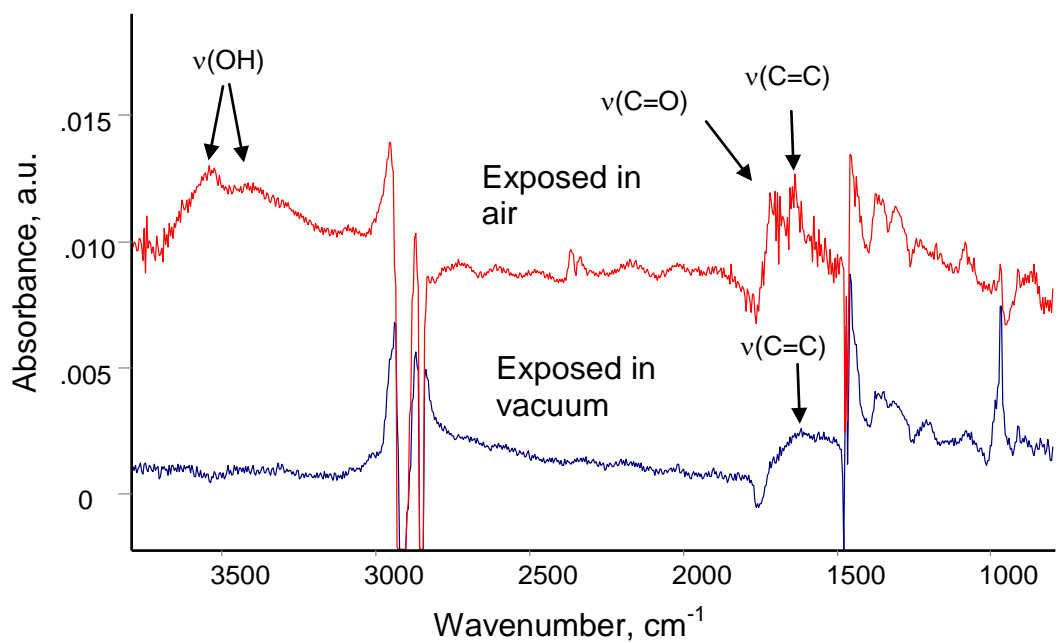

Fig.6. Differential FTIR ATR spectra of PIII treated UHMWPE sample exposed to air and exposed to vacuum after treatment. The PIII treatment time is 200 sec for both samples. The sample without air contact does not have oxygen containing groups in the surface layer. The spectra were recorded on single reflection hemispherical ATR Germanium crystal.

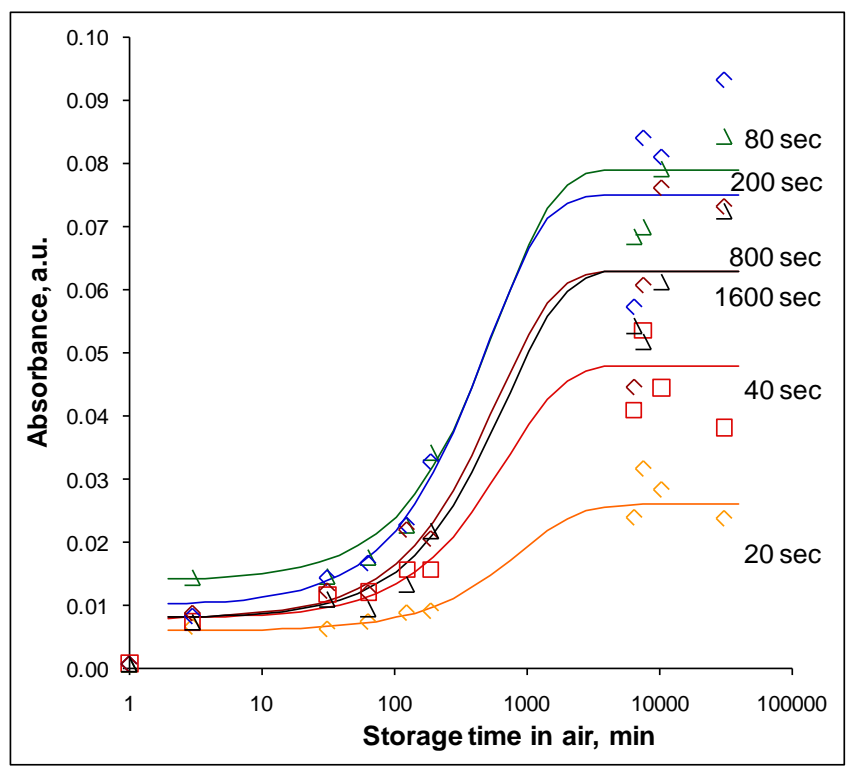

Fig.7. The ATR-FTIR absorbance intensity due to the 1712 cm$^{-1}$ carbonyl line $\nu$(C=O) normalized to the 2917 cm$^{-1}$ line of $\nu$(CH$_2$) stretch vibrations of UHMWPE as a function of storage time after PIII treatment for various times of PIII treatment.

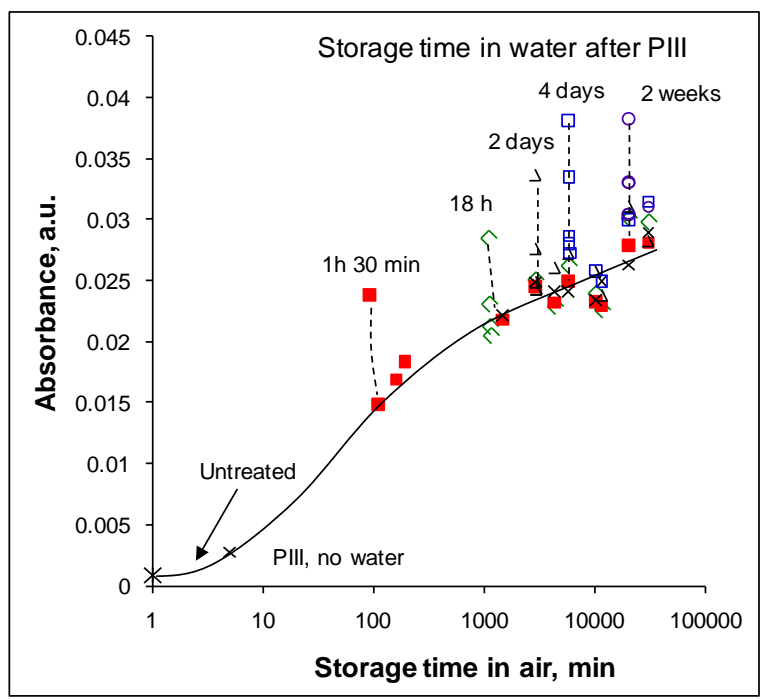

Fig.8. Normalized absorbance intensity due to hydroxyl groups (3400 cm$^{-1}$) in FTIR ATR spectra of UHMWPE as a function of time of storage after PIII treatment. The time of storage under water is shown annotated in the Figure. When the sample is taken from water, the adsorbed water in gives a contribution in the intensity of $\nu$(OH) line. After 30-40 mins the water evaporates.

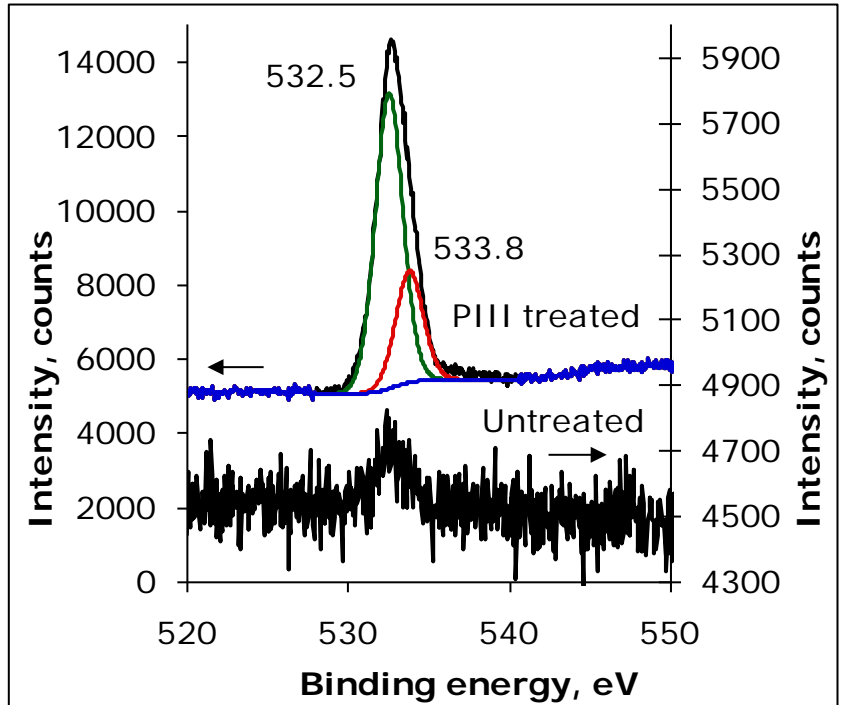

Fig.9a. $O_{1s}$ line of XPS fitted spectra of UHMWPE before and after PIII treatment with Ar ions of 20 keV energy and $10^{16}$ ions/cm$^2$ fluence. Untreated polyethylene has 532.5 eV low intensity line of oxygen in C=O bond. The PIII treated polyethylene has two $O_{1s}$ lines: 532.5 eV of C=O and N=O bonds and 533.8 eV of C-O and N-O bonds.

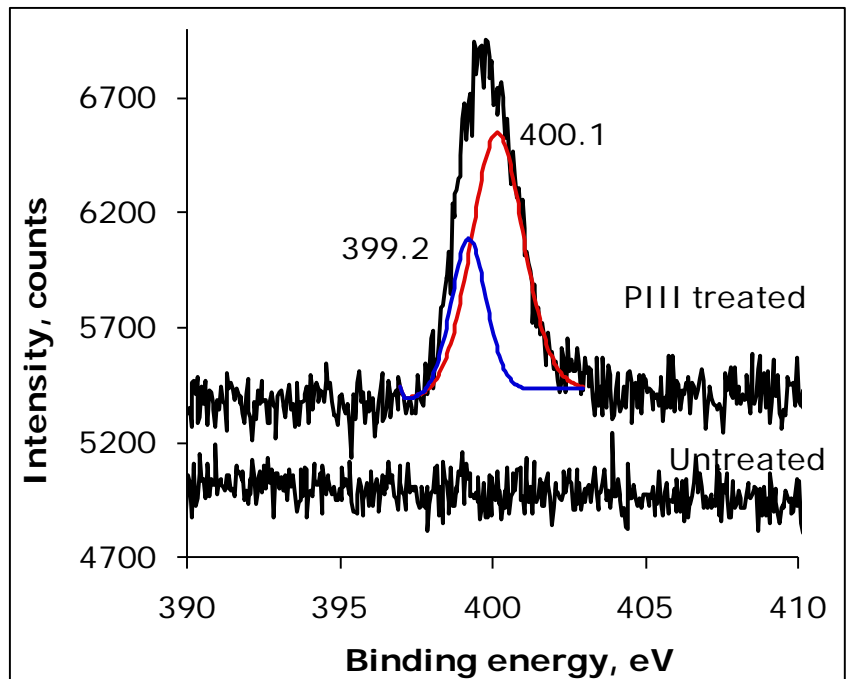

Fig.9b. $N_{1s}$ line of XPS fitted spectra of UHMWPE before and after PIII treatment with Ar ions of 20 keV energy and $10^{16}$ ions/cm$^2$ fluence. The untreated polyethylene does not contain nitrogen. The PIII treated polyethylene has two $N_{1s}$ lines: 399.2 eV of C=N and N=O bonds and 400.1 eV of C-N and N-O bonds.

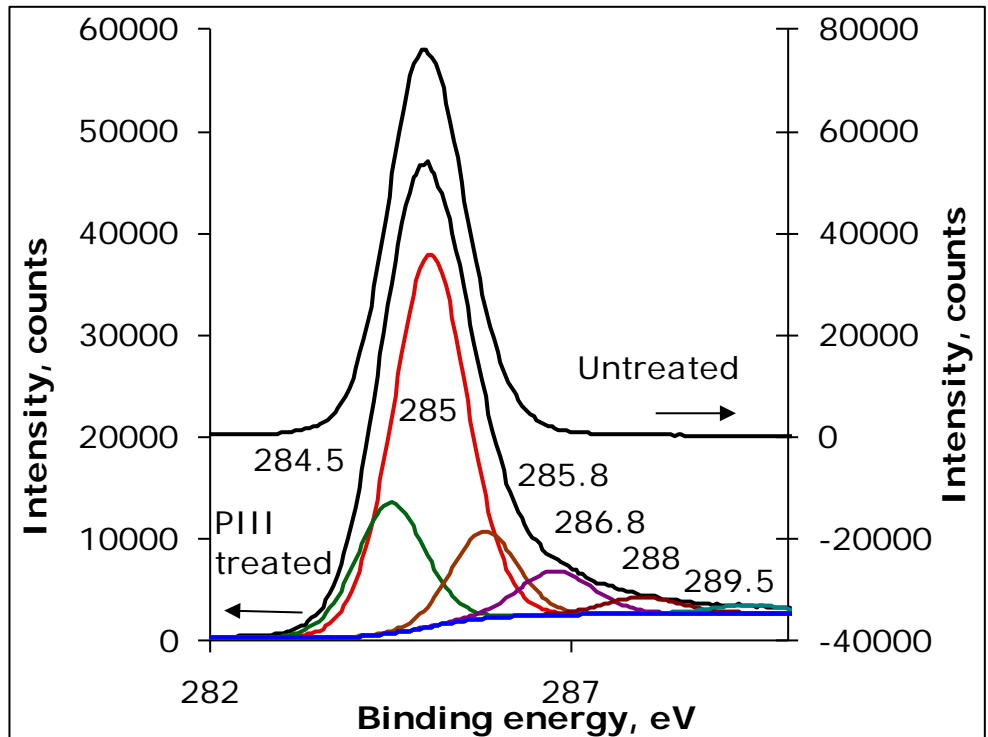

Fig.9c. $C_{1s}$ line of XPS fitted spectra of UHMWPE before and after PIII treatment with Ar ions of 20 keV energy and $10^{16}$ ions/cm$^2$ fluence. The untreated polyethylene has singlet $C_{1s}$ line at 285 eV. The PIII treated polyethylene has a number of $C_{1s}$ lines: 284.5 eV and 285 eV of hydrocarbon bonds, and 285.8, 286.8, 288 and 289.5 eV of carbon in various oxygen and nitrogen bonds.

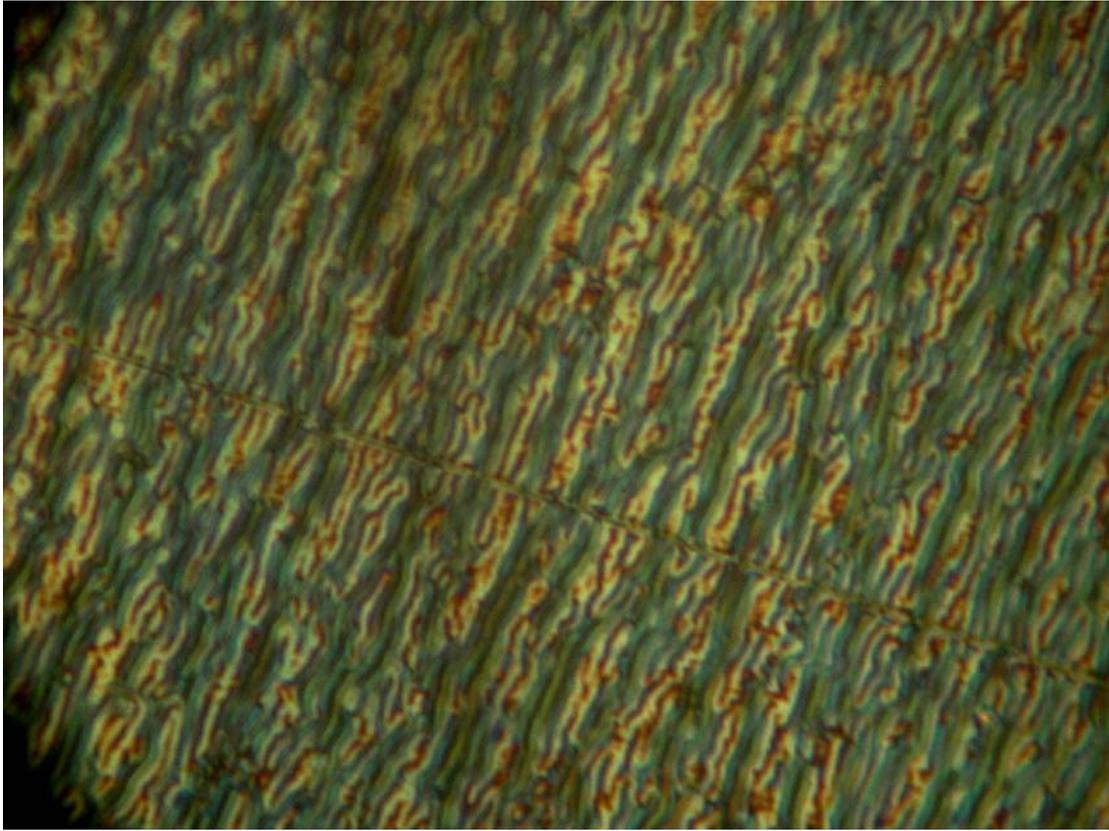

Fig.10a. Optical microphotos (size of 120 μm x 92 μm) of polyethylene after PIII treatment. The wave structure appears after PIII treatment while untreated film is smooth.

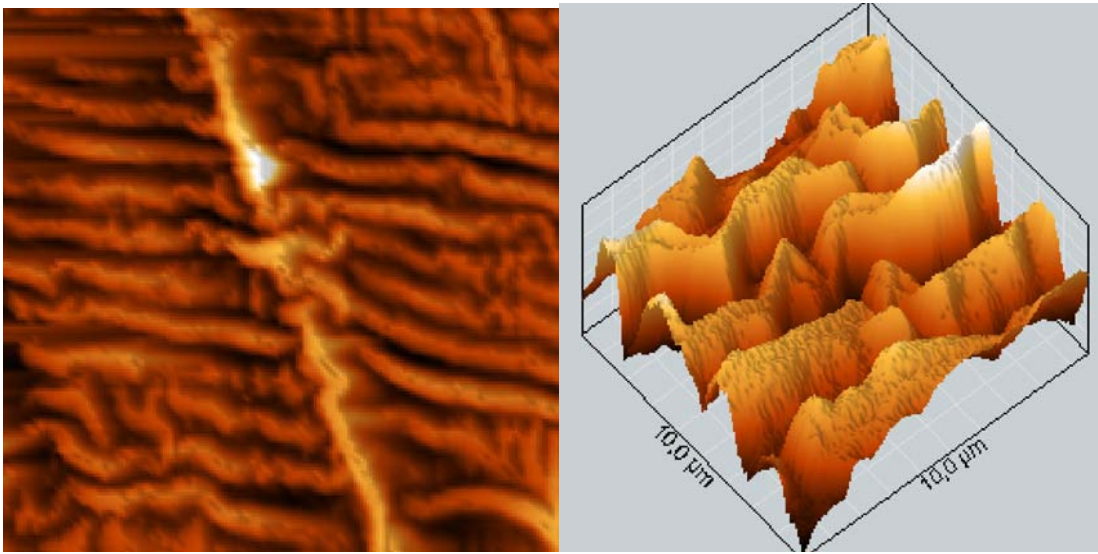

Fig.10b. AFM image of polyethylene surface after PIII. The lateral scale is 30 μm (left image) and 10 μm (right image). Vertical scale is 1 μm. The average wave's period is 2.3 μm.

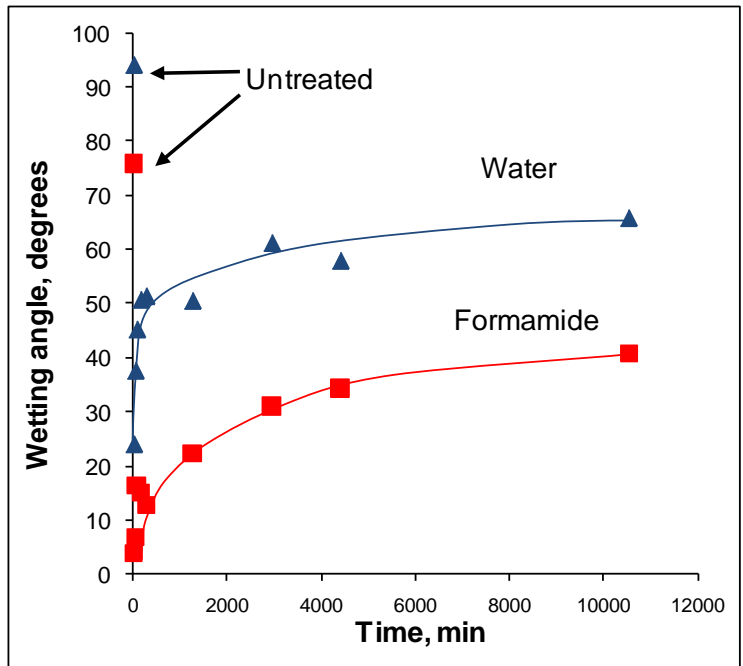

Fig.11. Water and formamide wetting angles of LDPE with time after PIII treatment. Marked signs correspond to untreated LDPE surface. Most recover of contact angle occurs shortly after PIII. However, the contact angle never comes back to initial.

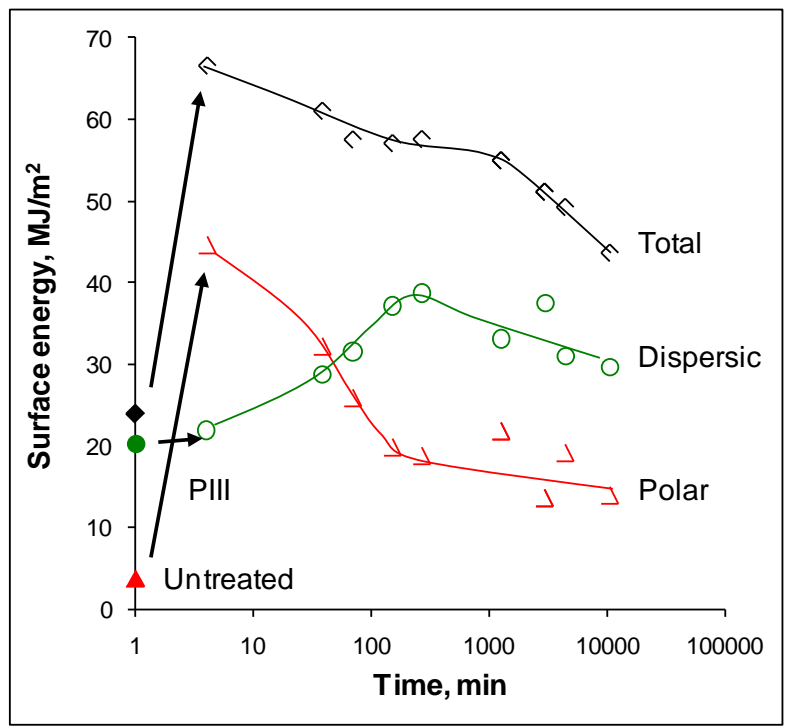

Fig.12. Total surface energy and its parts (polar and dispersic) of LDPE with time after PIII treatment. Full signs correspond to untreated LDPE surface. Empty signs correspond to PIII treated LDPE surface. Continuous lines correspond to equations of (1-3).

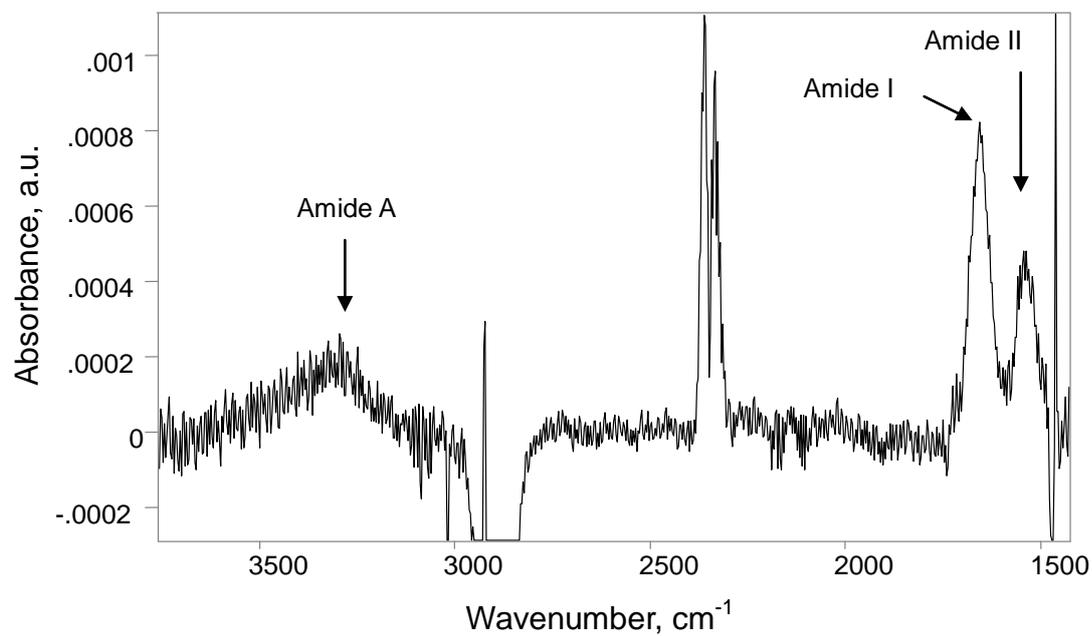

Fig.13. FTIR ATR spectra of attached HRP protein on LDPE. The spectra of LDPE is subtracted. Three main lines Amide A, I and II of protein are clear visible. The centre peak is $CO_2$ gas in FTIR spectrometer. The regions of polyethylene lines (2980-2820 cm$^{-1}$ and 1480-1450 cm$^{-1}$) are noisy after subtraction.

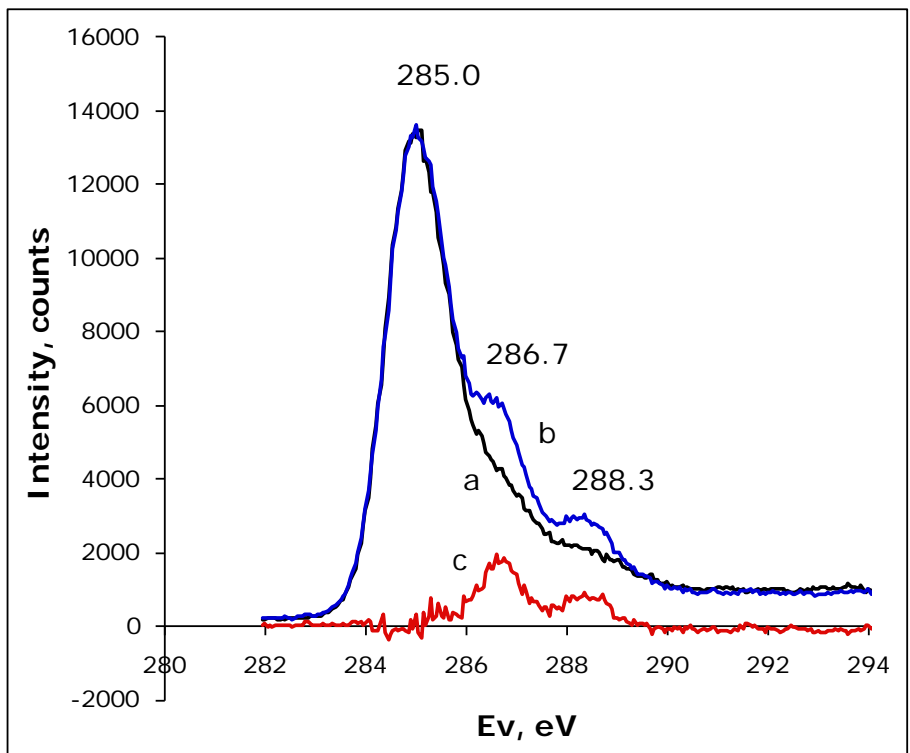

Fig.14a. XPS C$_{1s}$ line of N+ PIII treated UHMWPE without (a) and with (b) attached HRP protein and difference between these two spectra (c).

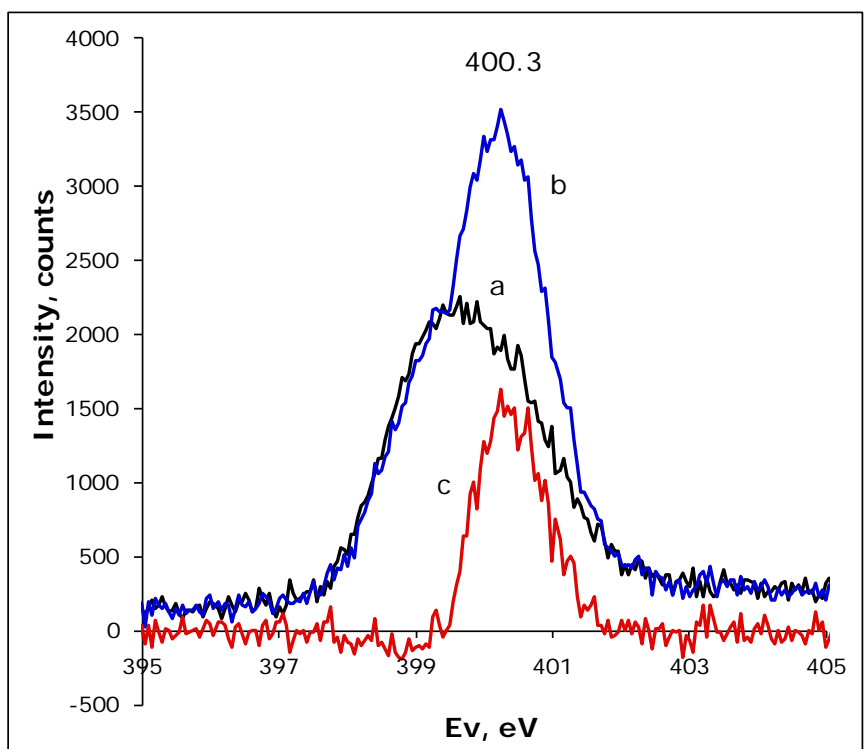

Fig.14b. XPS N$_{1s}$ line of N+ PIII treated UHMWPE without (a) and with (b) attached HRP protein and difference between these two spectra (c).

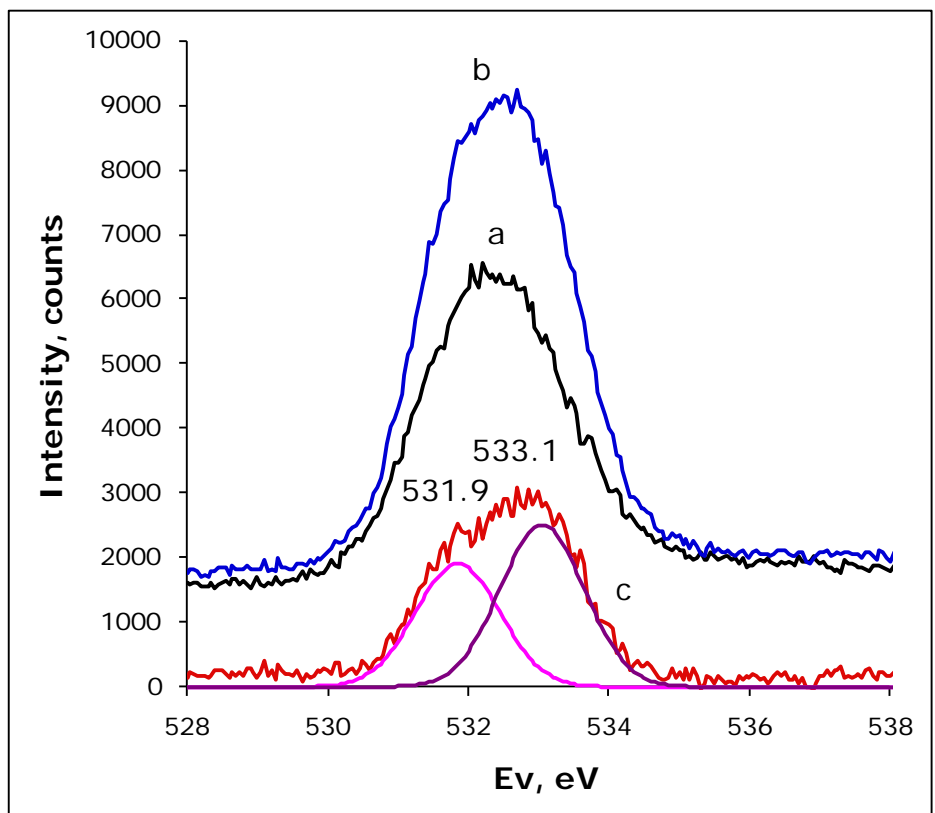

Fig.14c. XPS $O_{1s}$ line of N+ PIII treated UHMWPE without (a) and with (b) attached HRP protein and difference between these two spectra (c) with two fitted individual peaks.

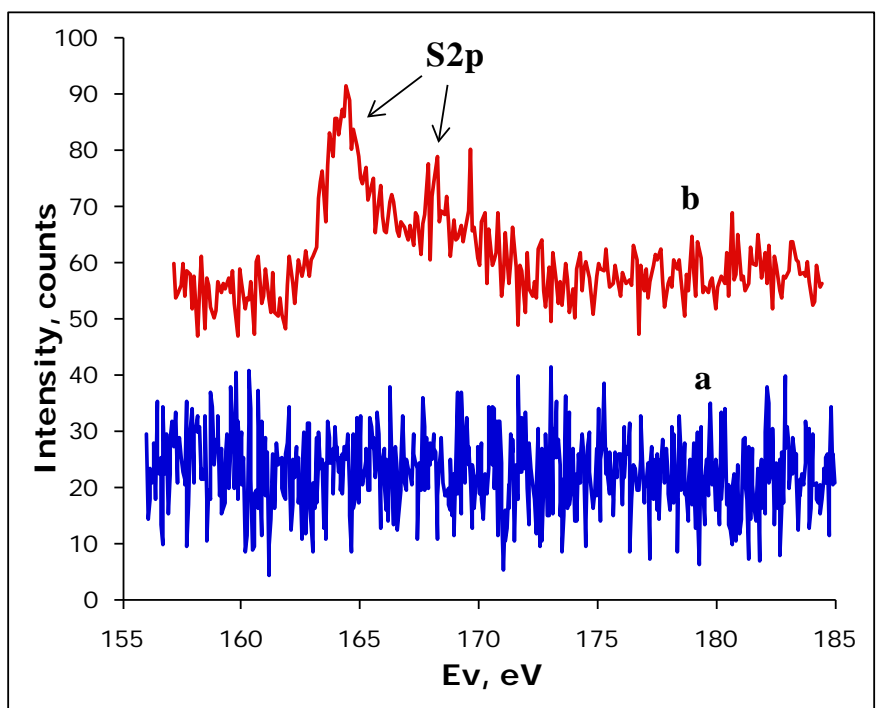

Fig.14d. XPS $S_{2p}$ lines of N+ PIII treated UHMWPE without (a) and with (b) attached HRP protein.

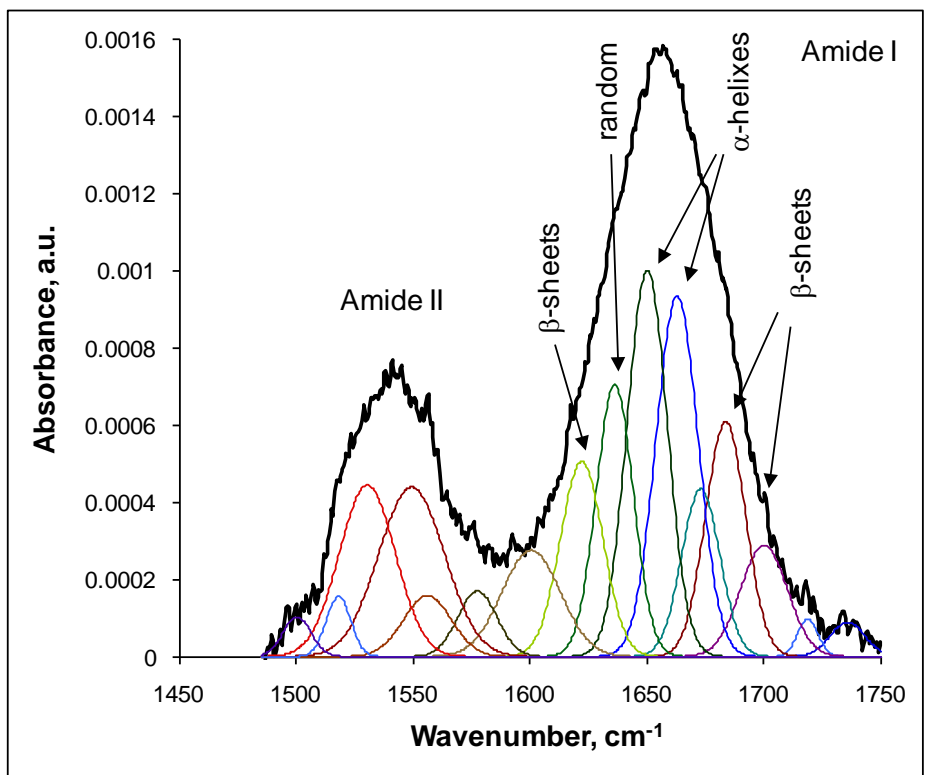

Fig.15. Fitting results of Amide I and II region in FTIR ATR spectra of dry attached HRP on PIII treated polyethylene.

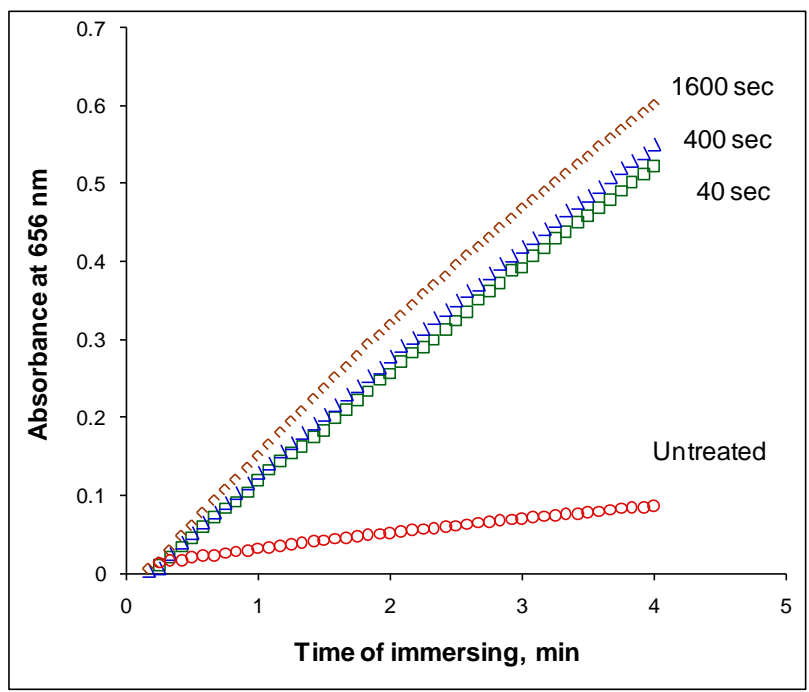

Fig.16a. Absorbance of TMB solution in dependence on the immersing time of UHMWPE surface with attached HRP. The curves of various PIII treatment time and the curve for untreated polyethylene are marked.

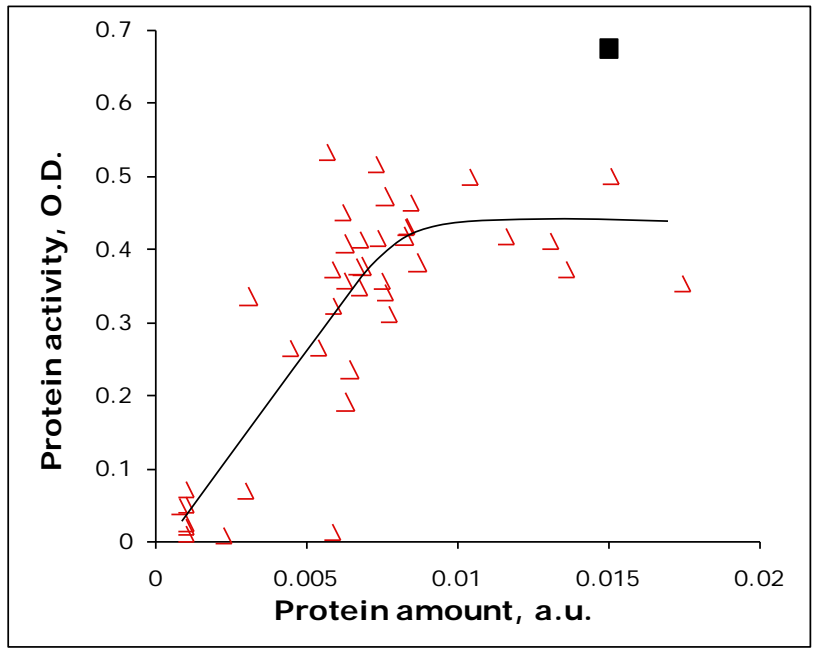

Fig.16b. Catalytic activity of attached HRP on PIII treated polyethylene in dependence of protein amount determined by FTIR ATR spectra. Full square is activity of HRP in solution with the concentration proportional to a complete monolayer of the protein on the surface.

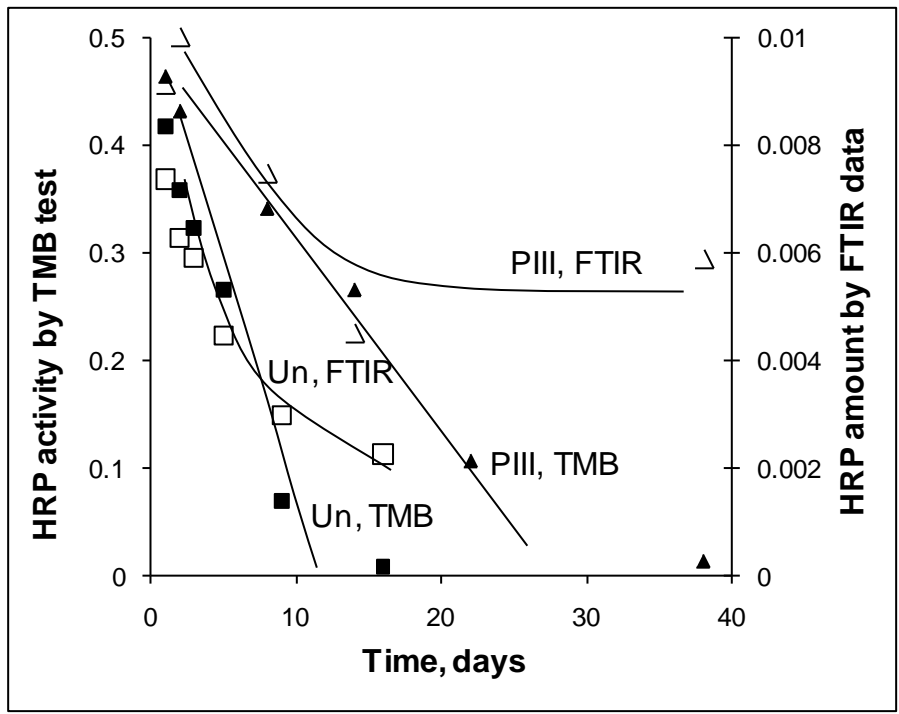

Fig.17. HRP activity (by TMB test data) and amount of HRP (by FTIR data) on untreated (Un) and PIII treated (PIII) UHMWPE surface with storing time in buffer solution at room temperature after protein attachment.

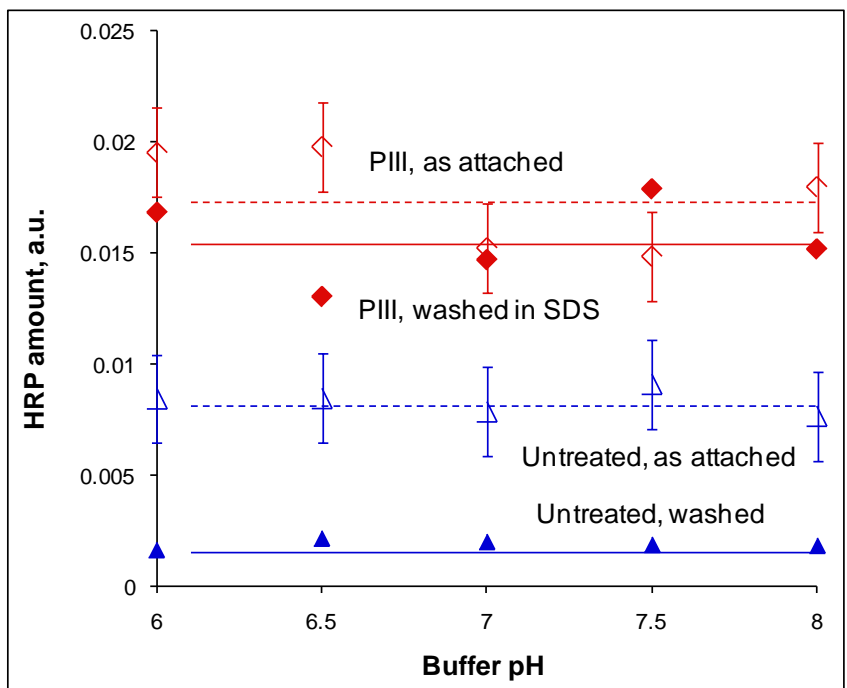

Fig.18a. Amount of attached HRP protein on untreated (blue) and PIII treated (red) polyethylene in dependence on pH of HRP solution before (empty signs) and after washing in SDS detergent during 1 hour at $70^0$ C (full signs). The amount of attached protein does not depend on pH of buffer solution.

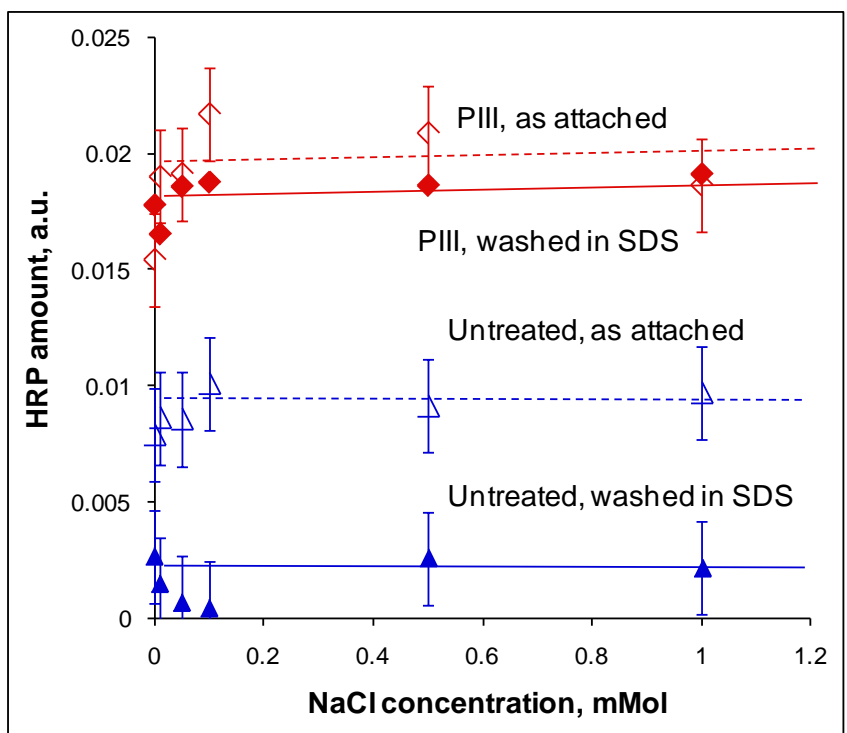

Fig.18b. Amount of attached HRP protein on untreated (blue) and PIII treated (red) polyethylene in dependence on salt NaCl concentration of HRP solution before (empty signs) and after washing in SDS detergent during 1 hour at $70^0$ C (full signs). The amount of attached protein does not depend on salt concentration.

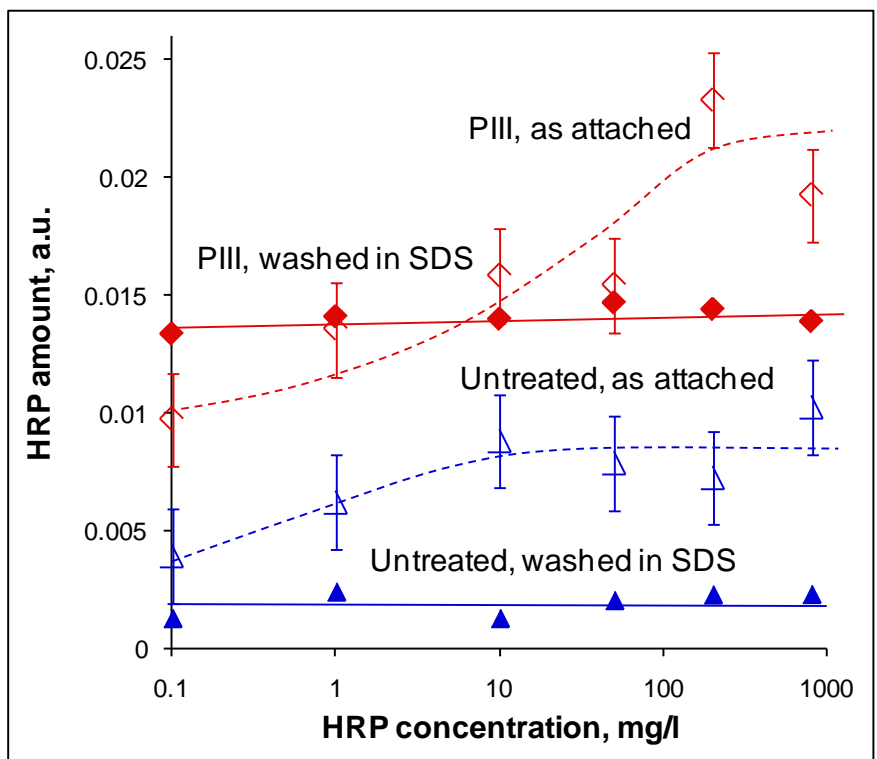

Fig.18c. Amount of attached HRP protein on untreated (blue) and PIII treated (red) polyethylene in dependence on concentration of HRP solution before (empty signs) and after washing in SDS detergent during 1 hour at $70^0$ C (full signs).

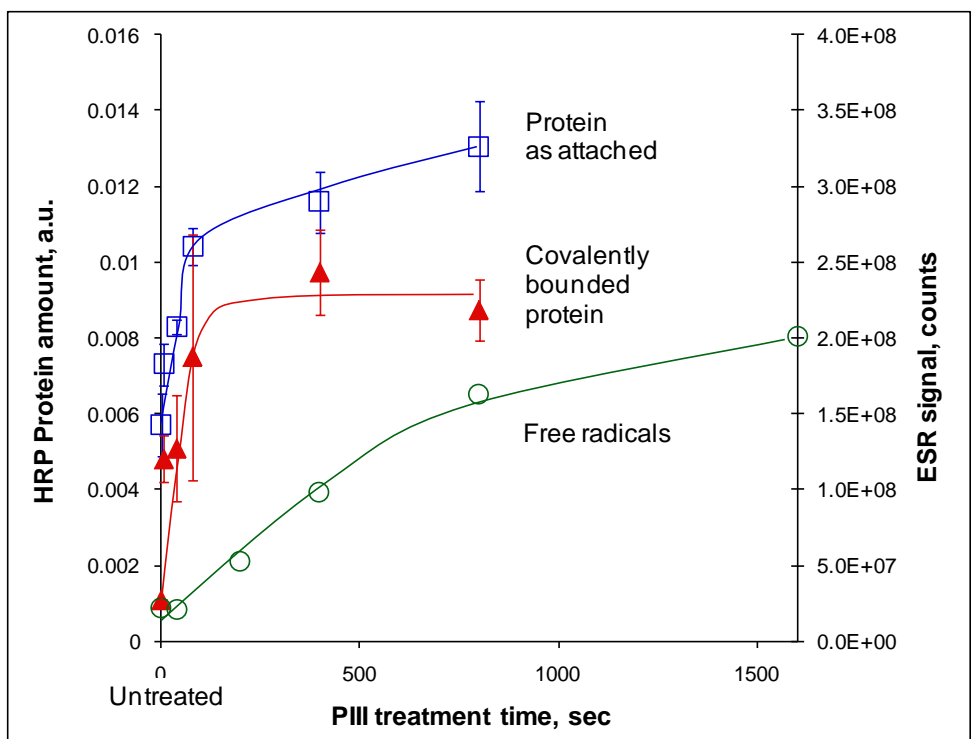

Fig.18d. Amount of attached HRP protein on polyethylene in dependence on PIII treatment time before (empty signs) and after washing in SDS detergent (full signs). The saturation of protein amount with PIII treatment time is observed for total protein amount and for amount of covalently attached protein. Low amount of free radicals is enough to achieve the saturation stage of protein attachment.

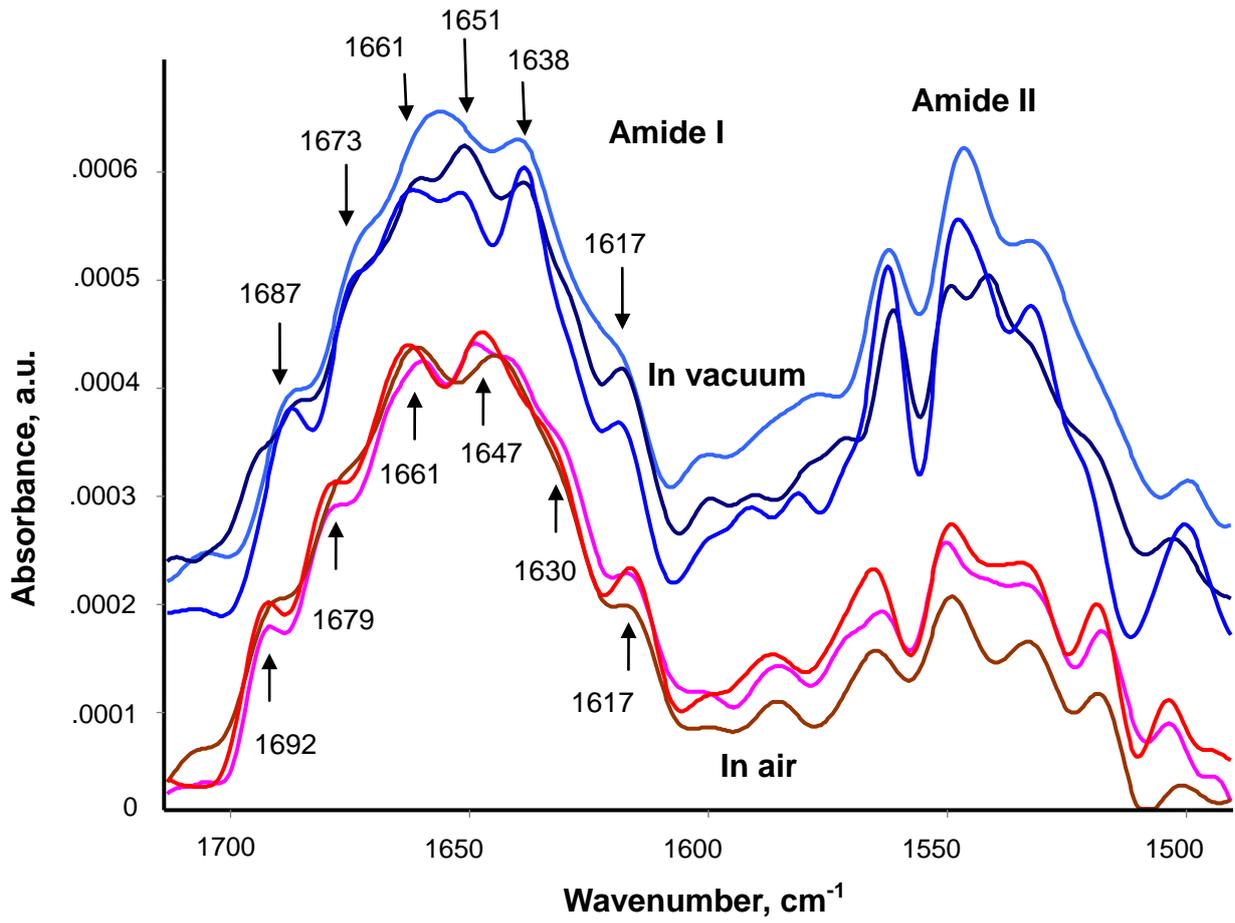

Fig.19. Self-deconvoluted FTIR ATR spectra of HRP attached on PIII treated LDPE exposed to air (three bottom spectra) and vacuum (three top spectra) for three different samples of each treatment. The spectrum to spectrum deviation for different samples is stronger for samples that had protein attached in vacuum (top spectra).

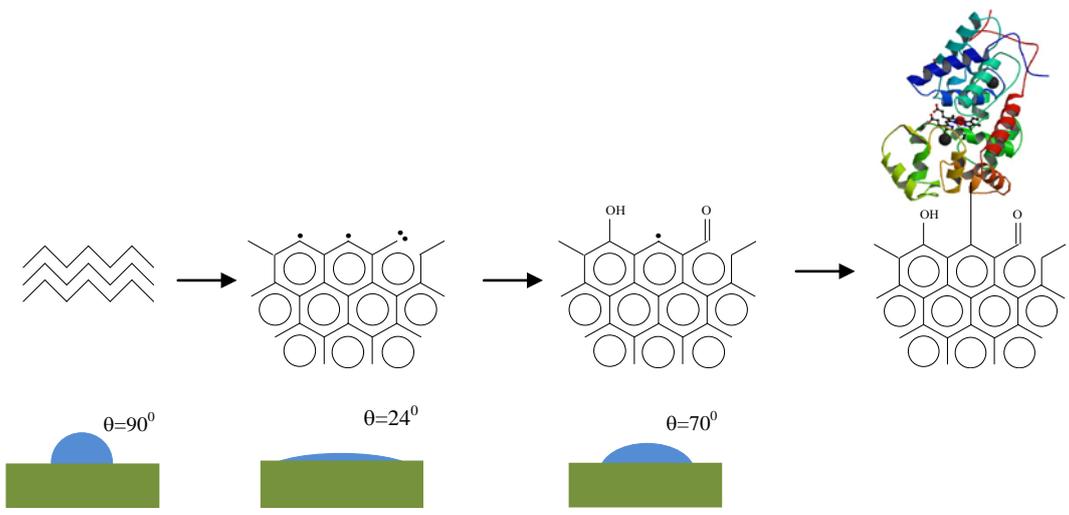

Fig. 20. Stages of protein attachment: untreated polyethylene with high contact angle; PIII ytreated polyethylene with high concentration of free radicals and low contact angle; aged PIII modified polyethylene with oxygen-containing groups and residual free radicals stabilised on π-electrons; protein is bounded to residual free radicals.